\documentclass[11pt]{article}

\usepackage[a4paper,margin=1in,top=1.1in,bottom=1.1in]{geometry}
\usepackage{amsmath,amssymb,amsthm}
\usepackage{graphicx}
\usepackage{booktabs}
\usepackage{multirow}
\usepackage{array}
\usepackage{xcolor}
\usepackage{hyperref}
\usepackage{url}
\usepackage{enumitem}
\usepackage{caption}
\usepackage{subcaption}
\usepackage{titlesec}
\usepackage{bm}
\usepackage{fancyhdr}
\usepackage{titletoc}
\usepackage{makecell}
\usepackage{multicol}
\usepackage[numbers,square,sort&compress]{natbib}

\hypersetup{
  colorlinks=true,
  linkcolor=blue!60!black,
  citecolor=blue!60!black,
  urlcolor=blue!60!black
}

\titleformat{\section}{\Large\bfseries}{\thesection.}{0.6em}{}
\titleformat{\subsection}{\large\bfseries}{\thesubsection.}{0.6em}{}
\titleformat{\subsubsection}{\normalsize\bfseries}{\thesubsubsection.}{0.5em}{}

\DeclareCaptionLabelFormat{barformat}{#1~#2~$|$}
\title{\vspace{-1.0cm}
{\bfseries\Huge TGR:}\\[8pt]
{\bfseries\LARGE Advancing Industrial Recommendation}\\[2pt]
{\bfseries\LARGE from Generative-Paradigm Ranking toward}\\[2pt]
{\bfseries\LARGE Unified Generation and Reasoning}}

\author{\Large TGR Team}
\date{}

\begin{document}
\maketitle
\thispagestyle{fancy}

\begingroup
\setlength{\parindent}{0pt}
\setlength{\parskip}{0.5em}
Industrial recommender systems have long followed a cascaded
architecture---retrieval, pre-ranking, ranking, and reranking---powered by
separately optimized Deep Learning Recommendation Models (DLRMs). This blueprint now
faces three structural challenges. First, fragmented DLRM architectures process high-cardinality, heterogeneous features through separately designed modules and exhibit no LLM-style scaling law: quality saturates as model capacity and compute grow. Second, stage-wise optimization and progressive candidate truncation fragment the decision
process, and pointwise item scoring ignores the inter-item dependencies and
positional effects that determine whole-slate quality. Third, co-occurrence-driven,
ID-based models lack the multimodal semantics, world knowledge, and deliberate
reasoning needed for cold-start, long-tail, and ambiguous-intent requests.
Together, these challenges are driving an industry-wide shift from the
cascaded toward the generative paradigm: unified, scalable sequence modeling
that unlocks predictable scaling, end-to-end generation that decides the
slate as a whole, and LLM-derived priors that supply world knowledge and
reasoning.
\par\noindent
To address these challenges across the heterogeneous business scenarios
within Tencent's PCG (Platform and Content Group), spanning diverse content
modalities, user populations, candidate-set sizes, and latency requirements, we develop
\textbf{TGR (Tencent Generative Recommendation)}, an industrial framework that advances
recommendation from conventional cascades toward the generative paradigm along three coupled directions. First, \textbf{TGR-GenRank} upgrades the ranking stage to
generative-paradigm ranking. Its production model, CCFormer, combines unified
feature tokenization and a scalable Transformer backbone with feature-field
separated cross attention, long-sequence subspace token mixing, and hierarchical sequence
compression, while retaining per-item multi-task scores for drop-in deployment. Second, \textbf{TGR-GenRec} explores end-to-end replacement of the
cascade's decision stages through two complementary generation paradigms.
Under next-token prediction (NTP), BARGE formalizes and bridges two
structural gaps in hierarchical semantic-ID generation---the loss of item
boundaries when multi-token items are flattened and semantic drift caused by
early prefix errors---through item context-aware attention, hierarchical
path reranking, and orthogonal dual-path decoding. Under next-slate
prediction (NSP), HiGR reformulates recommendation from item-by-item decoding
to whole-slate generation and co-designs prefix-structured semantic IDs for
controllable planning, a coarse-to-fine decoder that separates global slate
planning from local item generation, and listwise multi-objective alignment
over ranking fidelity, genuine user interest, and diversity. Third, \textbf{TGR-Reason} augments end-to-end generation with LLM reasoning
amortized through latent multi-step supervision. Offline, a LatentRec-trained
Think model materializes complete semantic-ID \emph{reason tokens}; Direct
Reasoning Injection uses them as hierarchy-aligned prompts to condition
online decoding without a request-time reasoning rollout.
\par\noindent
TGR is deployed at scale on Tencent production traffic, serving hundreds of
millions of users across surfaces including news, video, and web novels.
CCFormer outperforms HSTU, OneTrans, and STCA on public benchmarks and a
4-billion-sample production dataset, scales predictably at roughly half the
GFLOPs of HSTU, trains 2.21$\times$ faster, delivers statistically
significant lifts in all five A/B-tested production scenarios, and is fully launched in two of them (+3.57\% CTR in the video-recommendation scenario; +1.71\% advertising revenue in the advertising-ranking scenario). BARGE
outperforms OneRec by 10.2--16.9\% in Hit@5 across two industrial scenarios
and is fully rolled out as a generative retrieval channel with parameters,
beam width, and serving latency unchanged, delivering +0.60\% CTR and
+1.70\% total reading time---gains that grew larger and more stable after
full launch---and the highest CTR among the platform's representative retrieval channels.
HiGR improves offline slate quality by 15.9--21.3\% over matched-capacity
OneRec with a 5$\times$ inference speedup, serves whole slates below 50\,ms
P99 latency at roughly 60\% lower GPU demand than full-sequence decoding,
and consistently lifts engagement, traffic consumption, and commercial value
across three A/B-tested scenarios (up to +1.22\% Average Watch Time and
+1.73\% Average Video Views, +0.68\% CTR, +0.56\% advertising revenue).
Offline, TGR-Reason raises cold-start new-user Hit@1 by +477.8\%; online, it
delivers statistically significant gains of +1.75\% in Effective Consumption
Rate and +13.09\% in new-user Exposure-to-Conversion Rate.
\endgroup

\vspace{4pt}
\noindent\rule{\linewidth}{0.4pt}

{\small\tableofcontents}
\noindent\rule{\linewidth}{0.4pt}

\bigskip

\section{Introduction}
\label{sec:intro}

For a decade, industrial recommender systems have shared a common blueprint: a cascaded architecture of retrieval, pre-ranking, ranking, and reranking, in which the major decision stages are powered by separately optimized Deep Learning Recommendation Models (DLRMs) \citep{wide_deep_2016,youtube_dnn_2016,deepfm_2017,sasrec_2018,bert4rec_2019}. The blueprint is modular and operationally convenient, and it has carried industrial recommendation through years of continuous evolution. Yet it is now facing three structural challenges at once---challenges that, we argue, cannot be tuned away from inside the blueprint:

\begin{enumerate}[label=\textbf{(\roman*)},leftmargin=2.2em,itemsep=2pt,topsep=2pt]
\item \textbf{The ranking architecture no longer converts compute into quality.} A production DLRM processes high-cardinality, heterogeneous features through separately designed modules for feature extraction, feature interaction, and representation transformation \citep{hstu_2024}. This fragmented architecture is memory-bandwidth-bound, sustains low Model FLOPs Utilization, and exhibits no LLM-style scaling law: added parameters no longer buy quality, and feature interactions are bounded by what can be hand-engineered \citep{wukong_2024}. Generative-paradigm rankers offer the first credible break through this ceiling---by tokenizing user, item, context, and sequence features into a single token stream and letting a large Transformer do the crossing, HSTU scales sequential transducers into the trillion-parameter regime and ships $+12.4\%$ in online A/B at Meta \citep{hstu_2024}, while OneTrans lifts per-user GMV by $+5.68\%$ in production at strict latency budgets \citep{onetrans_2025}---but their self-attention cost grows quadratically with behavior-sequence length, so fine-grained cross-field interaction is routinely sacrificed, via sequence pre-compression or truncation, to contain cost.
\item \textbf{Stage-wise, item-wise decisions misalign with list-level experience.} Stage-wise optimization and progressive candidate truncation fragment the decision process, causing information loss and objective mismatch across the cascade. The fragmentation runs deeper than the pipeline: what a user actually experiences is an ordered \emph{slate} in a single display, whose quality hinges on the inter-item dependencies and positional effects that pointwise, item-in-isolation scoring cannot capture \citep{listcvae_2018}---and the direct generative remedy, decoding an $M$-item slate of $D$-token semantic IDs autoregressively, is too slow under real-time constraints.
\item \textbf{Pattern matching is not reasoning.} Conventional recommender models rely heavily on behavioral co-occurrence patterns learned from historical user--item interactions and encoded in ID-based representations. Their limited access to multimodal item semantics, world knowledge, and deliberate reasoning constrains the inference of nuanced or ambiguous user intents and generalization under sparse supervision---precisely the cold-start, long-tail, and ambiguous-intent regimes in which recommendation quality is decided. LLM-derived priors offer the natural remedy. In existing approaches, however, the reasoning is executed for every request or refresh---as chain-of-thought text \citep{onerec_think_2025} or as multi-step latent rollouts \citep{rearec_2025}---a cost that industrial latency budgets cannot absorb on the serving path and that offline precomputation only relocates.
\end{enumerate}

Catalyzed by the broader success of generative AI \citep{gpt4_2023,llama2_2023,deepseek_r1_2025,kimi_k2_2025}, the shift has unfolded along three axes: unified, scalable sequence modeling, carried by generative-paradigm rankers such as HSTU \citep{hstu_2024} and OneTrans \citep{onetrans_2025}; end-to-end generation, where Kuaishou's OneRec series \citep{onerec_v1_2025,onerec_tr_2025,onerec_v2_2025} replaces the retrieve--rank pipeline with a single generator at $0.1$B--$8$B parameters, while GPR \citep{gpr_2025} pursues a unified ``one-model paradigm'' for advertising; and LLM-derived reasoning, where OneRec-Think \citep{onerec_think_2025} adds in-text reasoning to OneRec, and OneReason \citep{onereason_2026} achieves a thinking mode that outperforms the non-thinking mode on Kuaishou business benchmarks. Together, these systems establish that the cascade can be revisited piece by piece or replaced outright; they are the systems TGR benchmarks against or is positioned alongside in Table~\ref{tab:landscape}.

\begin{table}[t]
\centering
\caption{TGR vs.\ representative industrial generative recommenders. ``Item-AR'' = item-level autoregressive decoding; ``Slate-AR'' = list-level autoregressive decoding; ``CoT'' = chain-of-thought reasoning; ``reason tokens'' = reasoning-enriched semantic IDs generated offline by a LatentRec-trained Think model (\S\ref{sec:reason_latent}); ``PRL'' = per-step reasoning loss.}
\label{tab:landscape}
\small
\setlength{\tabcolsep}{4pt}
\resizebox{\textwidth}{!}{%
\begin{tabular}{l|l|l|l|l|l}
\toprule
System & Stage replaced & Tokenizer & Decoder & Alignment & Reasoning \\
\midrule
HSTU \citep{hstu_2024}            & Retrieval / ranking & Sequence ID & Sequential transducer & Multi-task & --- \\
OneTrans \citep{onetrans_2025}    & Ranking             & Unified tokens & Causal Transformer & Pointwise & --- \\
OneRec-V1 \citep{onerec_v1_2025,onerec_tr_2025}  & Retrieval + ranking    & RQ-Kmeans   & Enc--Dec, Item-AR     & DPO       & --- \\
OneRec-V2 \citep{onerec_v2_2025}  & Retrieval + ranking    & RQ-Kmeans   & Lazy Decoder, Item-AR & GBPO + user feedback & --- \\
OneRec-Think \citep{onerec_think_2025} & Retrieval + ranking & RQ-Kmeans & Lazy Decoder, Item-AR & RL + In-text CoT & In-text CoT \\
OneReason \citep{onereason_2026}       & Retrieval + ranking & RQ-Kmeans & LLM decoder (Qwen3), Item-AR & CoT SFT + RL & In-text CoT (3-level) \\
GPR \citep{gpr_2025}              & Full pipeline (ads) & Unified rep.\ & HHD                & VAFT + HEPO (RL) & --- \\
PinRec \citep{pinrec_2025}        & Retrieval           & Multi-token & Multi-token AR        & Outcome-cond. & --- \\
\textbf{TGR-GenRank (CCFormer)}  & \textbf{Ranking}    & Unified tokens & Cross-field attn.\ + compr. & Multi-task & --- \\
\textbf{TGR-GenRec (BARGE)}         & \textbf{Retrieval} & OSQ-VAE (dual) & Dual-decoder, Item-AR + HPR & Aux.\ InfoNCE & --- \\
\textbf{TGR-GenRec (HiGR)}          & \textbf{Retrieval / Retrieval + ranking} & PCRQ-VAE & Hierarchical, Slate-AR & ORPO & --- \\
\textbf{TGR-Reason}                & \textbf{Retrieval / Retrieval + ranking} & Shared SID (PCRQ-VAE) & Slate-AR + reason tokens & NTP + PRL & Latent (offline) \\
\bottomrule
\end{tabular}}
\end{table}

\paragraph{Our setting.} Within Tencent's Platform and Content Group (PCG), recommendation runs across a deliberately heterogeneous portfolio of surfaces---news feeds, short videos, long videos, music, and web novels---which together serve hundreds of millions of daily active users, with candidate corpora of tens to hundreds of millions of items and tight per-stage P99 latency budgets in the tens of milliseconds.

\begin{figure}[t]
\centering
\includegraphics[width=0.95\textwidth]{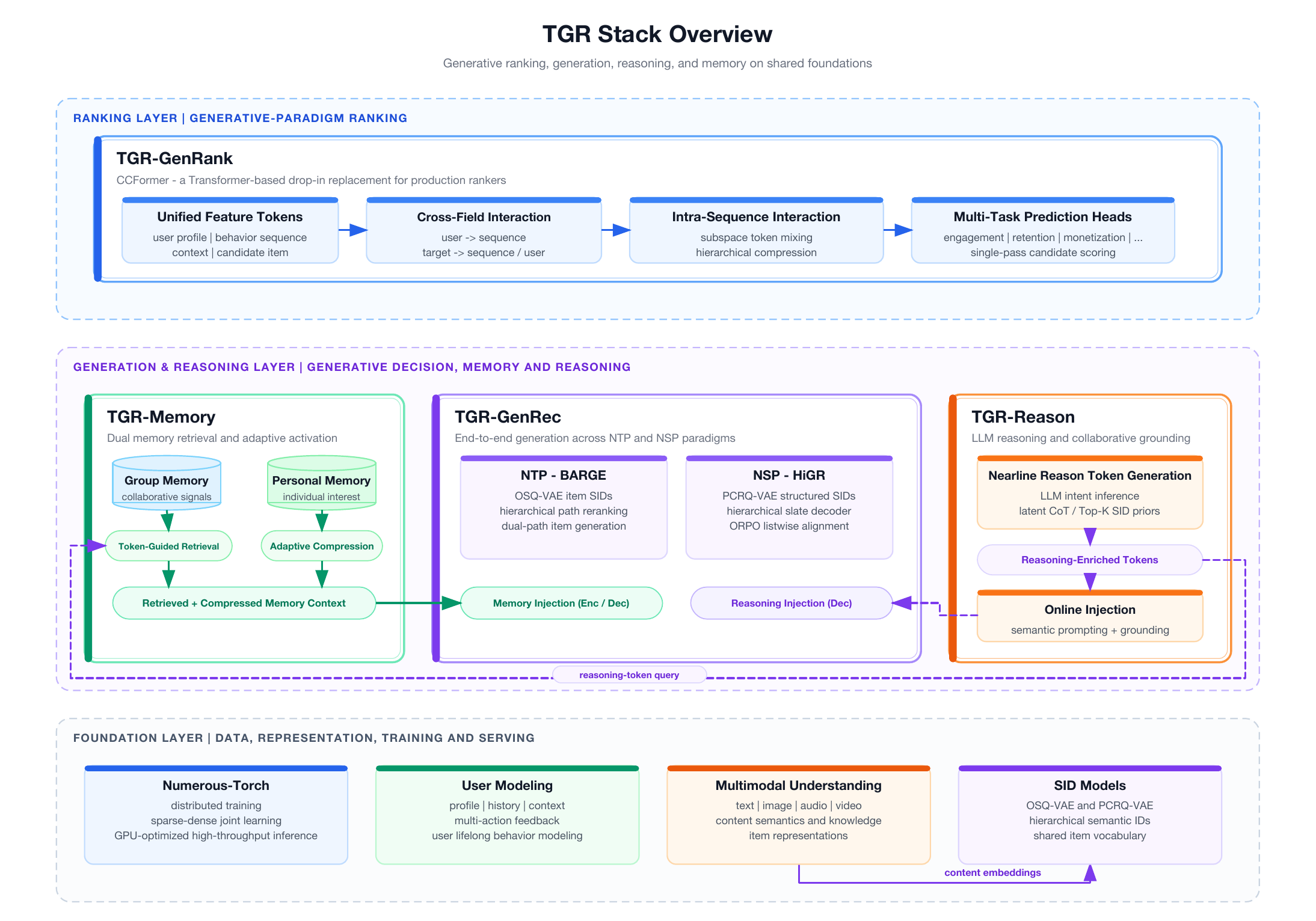}
\caption{Overall architecture of the TGR Stack. \emph{Top:} the ranking layer---TGR-GenRank (CCFormer) turns unified feature tokens into per-item multi-task scores through cross-field and intra-sequence interaction, with single-pass candidate scoring. \emph{Middle:} the generation and reasoning layer---TGR-GenRec generates the result list end to end under the NTP (BARGE) and NSP (HiGR) paradigms, with TGR-Reason injecting reasoning-enriched tokens and TGR-Memory supplying retrieved memory context (the Group Memory retrieval path is a design extension, excluded from the deployed configuration; \S\ref{sec:reason_inject}). \emph{Bottom:} the foundation layer---Numerous-Torch training and serving, user modeling, multimodal understanding, and the SID models (OSQ-VAE and PCRQ-VAE) shared across the stack.}
\label{fig:tgr_overview}
\end{figure}

\paragraph{The TGR Stack.} To address these challenges in this setting, we develop \textbf{TGR (Tencent Generative Recommendation)}, an industrial framework that advances recommendation from the cascaded blueprint toward the generative paradigm along three coupled directions---the \textbf{TGR Stack} (Figure~\ref{fig:tgr_overview}):

\begin{enumerate}[leftmargin=1.4em,itemsep=2pt,topsep=2pt]
\item \textbf{TGR-GenRank --- generative-paradigm ranking} (\S\ref{sec:ccformer}). Building on the HSTU \citep{hstu_2024} and OneTrans \citep{onetrans_2025} line of work, TGR-GenRank keeps the inherited cascade in place and replaces the production rankers with \textbf{CCFormer} \citep{ccformer_2027}, an efficient Transformer backbone that unifies cross-field feature interaction and compressed long-sequence modeling: feature-field separated cross attention lets user, behavior-sequence, and target-item tokens interact along semantically directed flows, while subspace token mixing and hierarchical sequence compression make long-sequence modeling subquadratic without giving up access to the full history. Trained with unified multi-task supervision on LLM-style infrastructure, CCFormer is deployed as a drop-in replacement for the production ranker in two scenarios. Note that, although TGR-GenRank adopts the generative paradigm's tokenized representation, Transformer backbone, and scaling recipe, the ranking output remains per-item scores---it \emph{borrows} the generative paradigm and is therefore best understood as ``generative-paradigm ranking'', not yet ``end-to-end generation''.
\item \textbf{TGR-GenRec --- unified end-to-end generative recommendation} (\S\ref{sec:higr}). The direction that breaks out of the cascade entirely: its goal is to replace retrieval, ranking, and reranking with a single generative model that decides the result list end to end, removing per-stage information loss and objective drift. Aligned with the one-model direction of OneRec at Kuaishou, TGR-GenRec currently ships one production model per generation paradigm (\S\ref{sec:genrec_overview}): \textbf{BARGE} \citep{barge_2027}, which structurally adapts the next-token-prediction (single-item) generation mechanism to the recommendation task through item context-aware attention, hierarchical path reranking, and orthogonal dual-path decoding; and \textbf{HiGR} \citep{higr_2026}, which moves up to whole-slate generation---differentiating from OneRec-V1/V2 by emitting the entire \emph{slate} as one structured object rather than decoding items one at a time---through Prefix-Contrastive RQ-VAE tokenization (PCRQ-VAE), a Hierarchical Slate Decoder (HSD), and ORPO-based listwise multi-objective alignment.
\item \textbf{TGR-Reason --- reasoning-augmented generative recommendation} (\S\ref{sec:think}). Among the possible carriers for connecting LLM reasoning with a generative recommender---chain-of-thought text, auxiliary generated features, latent reasoning states, and reasoning-enriched semantic IDs (\S\ref{sec:reason_taxonomy})---TGR-Reason adopts the last: a Think model trained with LatentRec internalizes latent reasoning at training time and periodically exports \emph{reason tokens} offline, which are injected into the online generator (and can further query Group Memory) without adding any reasoning rollout to the request path. In the FM4RecSys 3-paradigm taxonomy (Feature-Based $\to$ Generative $\to$ Agentic) \citep{fm4recsys_2025}, TGR-Reason is best understood as the on-ramp from \emph{Generative} toward \emph{Agentic}: it keeps a single generator but equips it with deliberate reasoning, while leaving room for tool use, multi-turn dialogue, and autonomous agents---hallmarks of the fully agentic paradigm---in later iterations.
\end{enumerate}

\paragraph{Contributions.} All three directions are deployed on our production traffic and validated in online A/B tests; this report consolidates the design decisions and the offline, scaling, ablation, and online evidence behind each. Its contributions are:

\begin{enumerate}[leftmargin=1.4em,itemsep=2pt,topsep=2pt]
\item We describe an industrially deployed generative recommendation stack (TGR) spanning three coupled directions---TGR-GenRank, TGR-GenRec, and TGR-Reason---under a common production discipline, with a semantic-ID tokenizer family, user/context encoder, and post-training pipeline shared across the generative-recommendation models. TGR currently serves hundreds of millions of users across multiple Tencent surfaces.
\item We provide a detailed technical description of \textbf{TGR-GenRank}'s production model \textbf{CCFormer}---feature-field separated cross attention, long-sequence subspace token mixing, and hierarchical sequence compression---with an emphasis on its industrial training and serving recipe (mixed-precision training, INT8 + double-hashing sparse-parameter compression, and single-pass parallel candidate scoring on the Numerous-Torch infrastructure), validated across five production scenarios, two of which are fully launched.
\item We provide a detailed technical description of \textbf{TGR-GenRec}, covering one production model per generation paradigm: BARGE, which structurally adapts next-token-prediction SID generation via item context-aware attention, hierarchical path reranking, and dual-path decoding; and HiGR, which realizes whole-slate generation through its tokenizer (PCRQ-VAE), coarse-to-fine decoder (HSD), and ORPO-based listwise multi-objective alignment---together with offline and online results for both.
\item We present the design and deployment of \textbf{TGR-Reason}: a LatentRec-trained offline Think model that amortizes reasoning at training time and exports reasoning-enriched semantic-ID priors (\emph{reason tokens}), injected into the production generator at zero request-path reasoning cost. On one of our commercial content platforms it delivers consistent offline gains, strongest for cold-start new users, and statistically significant online A/B lifts.
\end{enumerate}
The remainder of this report is organized as follows. We first introduce the TGR Stack and shared components (\S\ref{sec:tgr_stack}). We then describe TGR-GenRank (\S\ref{sec:ccformer}); the TGR-GenRec direction (\S\ref{sec:higr}), whose two production models are presented as independent, parallel works---BARGE (\S\ref{sec:barge}) and HiGR (\S\ref{sec:higr_model}); and TGR-Reason (\S\ref{sec:think}). Each model chapter reports its own deployment and online A/B results in place; Section~\ref{sec:conclusion} concludes with limitations and future directions. The author list is provided in Appendix~\ref{app:contributions}, a notation table in Appendix~\ref{app:notation}, consolidated default configurations in Appendix~\ref{app:configs}, and complexity analyses in Appendix~\ref{app:complexity}.

\section{The TGR Stack: Design Principles and Shared Foundations}
\label{sec:tgr_stack}

This section describes the design principles that govern TGR as a whole and the components that are shared across TGR-GenRank, TGR-GenRec, and TGR-Reason. Figure~\ref{fig:tgr_overview} shows how the pieces fit together: a ranking layer that upgrades the cascade in place, a generation-and-reasoning layer that replaces it end to end, and a foundation layer---tokenizers, encoders, and training infrastructure---on which both build.

\subsection{Design Principles}
\label{sec:overview_principles}

Three principles guide the entire TGR Stack:

\paragraph{(P1) Replace the cascade piece by piece, each piece industrial-grade.}
A useful generative recommender is one that can replace a heavily tuned discriminative module \emph{at full production quality} from day one. Each TGR model is therefore deployed as a drop-in replacement for a specific stage, behind the existing safety, dedup, and business-rule filters, and is held to two bars: (i)~match or exceed the offline metrics of the incumbent, and (ii)~stay within the same online latency and GPU budgets. This discipline recurs throughout the report: CCFormer ships as a drop-in ranker on the production feature space (\S\ref{sec:genrank_infra}), BARGE was deployed under a hard no-increase constraint on parameters, beam width, and serving latency (\S\ref{sec:barge}), and HiGR serves whole-slate generation within a strict latency-and-GPU envelope (\S\ref{sec:higr-efficiency}).

\paragraph{(P2) Share structure across the stack.}
Generative recommendation works at scale only if tokenizer, user encoder, and post-training infrastructure can be amortized across models. TGR therefore enforces a single semantic-ID (SID) tokenizer family (PCRQ-VAE for slate generation and the orthogonal dual-channel OSQ-VAE for NTP generation; \S\ref{sec:higr_pcrqvae}, \S\ref{sec:barge_method}), a common user/context encoding interface, and a common post-training discipline. Sharing the SID space is what lets TGR-Reason's reason tokens be injected natively into TGR-GenRec's decoder (\S\ref{sec:reason_inject}), and what will allow a future generative head for TGR-GenRank to be bootstrapped from TGR-GenRec's fine-grained item generator.

\paragraph{(P3) Optimize for the list, not the item.}
The unit of user experience on our surfaces is the displayed list, so the stack-level objective is defined over whole slates rather than isolated items. Each stage honors that objective at the level its paradigm permits: online, every model is gated on list- and session-level business outcomes; the slate-generation line trains and aligns directly at the list level, integrating ranking fidelity, genuine user interest, and diversity in a single listwise ORPO loss over slate-level preference pairs (\S\ref{sec:higr-method}) instead of fusing per-item rewards by hand; and the NTP line and the ranking stage supply the item-level fidelity that the list objective consumes (\S\ref{sec:genrec_overview}). This list-first commitment is consistent with the evidence that reward design often matters more than raw scale \citep{onerec_v2_2025} and with the emphasis on session-level objectives in scaled sequential transducers \citep{hstu_2024}.

\subsection{Problem Formulation}

Let $\mathcal{U}$ denote the set of users and $\mathcal{V}$ the set of items. For each user $u \in \mathcal{U}$, we observe a chronological interaction history $S^u = (v^u_1, v^u_2, \dots, v^u_n)$ together with a set of user, context, and request-level features $x^u$. On each request, the system returns an ordered slate of $M$ items
\[
O^u = (\hat{v}^u_1, \hat{v}^u_2, \dots, \hat{v}^u_M),
\]
optimized for holistic list-level user feedback (e.g., watch time, app stay time, completion, satisfaction) rather than for per-item scores in isolation.

TGR pursues this objective in two deployment regimes. \emph{Within the cascade}, the slate is still assembled by the downstream stages, and TGR-GenRank upgrades the ranker that drives them: CCFormer consumes the production feature space and emits per-item multi-task scores for all candidates of a request in a single pass (\S\ref{sec:ccformer}). \emph{End to end} (TGR-GenRec and TGR-Reason), a generative Transformer $\mathcal{F}_\theta$ decodes the slate directly as a sequence of semantic-ID tokens,
\[
O^u = \mathcal{F}_{\theta}(u, S^u, x^u),
\]
with TGR-Reason conditioning the same decoding on reasoning-enriched priors (\S\ref{sec:higr}, \S\ref{sec:think}). The full notation is summarized in Appendix~\ref{app:notation}.

\subsection{Shared Components}

\paragraph{Semantic-ID tokenizers.} The slate-generation models (HiGR and TGR-Reason) share the same tokenizer, PCRQ-VAE (\S\ref{sec:higr_pcrqvae}). BARGE, under the NTP paradigm, uses the orthogonal dual-channel OSQ-VAE (\S\ref{sec:barge_method}). In either case, each item $v$ is mapped to a hierarchical SID sequence $(s^1, s^2, \dots, s^D)$ in which high-level prefixes encode coarse semantic and collaborative structure and the terminal layer disambiguates individual items. TGR-GenRank sits outside this machinery: its drop-in constraint requires consuming the production ranker's existing feature space, so CCFormer tokenizes raw user, behavior, and target feature fields directly instead of semantic IDs (\S\ref{sec:ccformer}).

\paragraph{User/context encoding.} The generative-recommendation models adopt a common encoder--decoder interface: a Transformer-based encoder maps $(u, S^u, x^u)$ to a contextualized memory $C$ that serves as the key/value context for cross-attention in the decoders. TGR-Reason reuses TGR-GenRec's encoder--decoder backbone directly (\S\ref{sec:reason_inject}), so reasoning and memory signals enter generation without any representation conversion.

\paragraph{Training and serving infrastructure.} TGR is trained on \emph{Numerous-Torch} (\S\ref{sec:genrank_infra}), Tencent's Megatron-style 3D-parallel trainer with cross-GPU embedding caching. Default deployments run on clusters of NVIDIA H20 nodes with NVLink intra-node and 400\,Gbps RDMA inter-node. We use BFloat16 mixed precision, ZeRO-1 sharding for dense parameters, gradient checkpointing on the deepest layers, and TensorRT-style fused kernels for the generative decoder. Online serving runs on NVIDIA L20 GPUs with KV-caching enabled by default.

\section{TGR-GenRank: Generative-Paradigm Ranking}
\label{sec:ccformer}

\textbf{TGR-GenRank} is the TGR Stack's generative-paradigm ranking direction: it keeps the cascade in place but replaces the production ranker itself with a Transformer that borrows the generative paradigm---unified tokenized inputs, an LLM-style backbone with predictable scaling behavior, and LLM-style training infrastructure---while still emitting per-item multi-task scores. Its production model is \textbf{CCFormer} \citep{ccformer_2027}, an efficient Transformer backbone that unifies \emph{cross-field feature interaction} and \emph{compressed long-sequence modeling}. CCFormer confronts the central deployment obstacle of the HSTU / OneTrans line of work \citep{hstu_2024,onetrans_2025}: attention-based rankers unlock scaling laws, but quadratic self-attention over long behavior sequences is unaffordable under industrial latency and resource budgets. By resolving this tension architecturally rather than by truncating or pre-compressing the sequence, CCFormer is simultaneously \emph{more accurate and cheaper} than its attention-based predecessors: it trains $2.21\times$ faster than a strong HSTU baseline at higher AUC and GAUC. These offline gains translate into consistent online gains: CCFormer has been validated through online A/B experiments across \emph{five} recommendation ranking scenarios at Tencent, delivering statistically significant lifts on the core business metrics in each scenario.

Consistent with the industrial focus of this report, we keep the architectural description compact (\S\ref{sec:genrank_arch}) and devote most of the section to the training and serving recipe (\S\ref{sec:genrank_infra}) and the experimental evidence---offline comparisons, scaling behavior, ablations, and online A/B results (\S\ref{sec:genrank_offline}--\S\ref{sec:genrank_online}).

\subsection{Motivation}
\label{sec:genrank_motivation}

Industrial rankers differ primarily in \emph{how they realize feature interaction}: over which tokens, at what asymptotic cost, and how much of the user history is retained. Under this view, ranking architectures deployed at scale fall into three families, each with a characteristic limitation.

\paragraph{Conventional DLRMs.} The incumbent paradigm embeds sparse categorical features, concatenates them with dense features, and applies hand-designed crossing modules: explicit low-order crossing with deep networks \citep{deepfm_2017}, target-aware attention pooling \citep{din_2018}, or memory-augmented compression of long histories \citep{mimn_2019}. Such models are cheap to serve but treat behaviors as aggregated or short-range features rather than as sequential evidence, so pooled representations discard fine-grained signals and capacity is spent on static feature pairs instead of history--target interaction. They also saturate with added capacity, offering no usable scaling law.

\paragraph{Attention-based sequential rankers.} The second family models behaviors token by token with self-attention \citep{sasrec_2018}, and its industrial variants jointly attend over behavior sequences, item features, and user profiles in a single scalable backbone \citep{hstu_2024,onetrans_2025,stca_2026}. These models adaptively identify informative historical interactions and exhibit predictable scaling laws in \emph{both} sequence length and model capacity---the only family that scales along both axes. Their obstacle is cost: self-attention is quadratic in sequence length in both computation and memory, which is prohibitive when scoring massive traffic under latency budgets of tens of milliseconds. Two lossy remedies are common in production. \emph{Compress-then-interact} methods condense the sequence before feature interaction, sacrificing fine-grained signals and weakening history--target interaction. \emph{Retrieve-or-truncate} methods (the SIM / TWIN lineage \citep{sim_2020,twin_2023}) retain token-level modeling only within a target-relevant subsequence, discarding long-term interests while still paying full attention inside the retained window. Efficient-attention variants, in turn, optimize mainly the sequence-modeling term and remain weak at the cross-field interaction on which ranking depends.

\paragraph{Token-mixing rankers.} A third family sequentializes the feature space into unified tokens and exchanges information across them with lightweight mixing operators derived from MLP-based alternatives to Vision Transformers, avoiding quadratic self-attention. RankMixer \citep{rankmixer_2025} combines multi-head token mixing with per-token FFNs to model heterogeneous feature subspaces at near-linear cost, recovering predictable \emph{capacity} scaling on hardware-efficient primitives. Their limitation is scope: these operators target a small set of unified feature tokens, not behavior sequences of hundreds to thousands of items. Long-term interest must still be summarized into few tokens before mixing, reintroducing the information loss of the first family.

\paragraph{Our approach.} The three families point to one unmet requirement: \emph{sufficient and efficient feature interaction over full-length behavior sequences}. Insufficient sequence-level crossing weakens fine-grained preference discovery, whereas exhaustive self-attention provides the crossing at a cost that is deployable only by discarding history. CCFormer therefore decouples feature interaction into two orthogonal components and makes each subquadratic by construction: \emph{cross-field} interaction among user, sequence, and target fields, realized by directed field-separated cross attention that computes only the interaction directions ranking requires; and \emph{intra-sequence} interaction among behavior tokens, realized by extending token mixing from short unified token sets to the full sequence, combined with hierarchical compression that gives deeper blocks whole-history coverage at geometrically decreasing cost. CCFormer thus inherits the scaling behavior of the attention family and the efficiency of the token-mixing family without retrieving or truncating the sequence in advance.

\subsection{CCFormer Architecture in Brief}
\label{sec:genrank_arch}

\begin{figure}[t]
\centering
\includegraphics[width=0.97\textwidth]{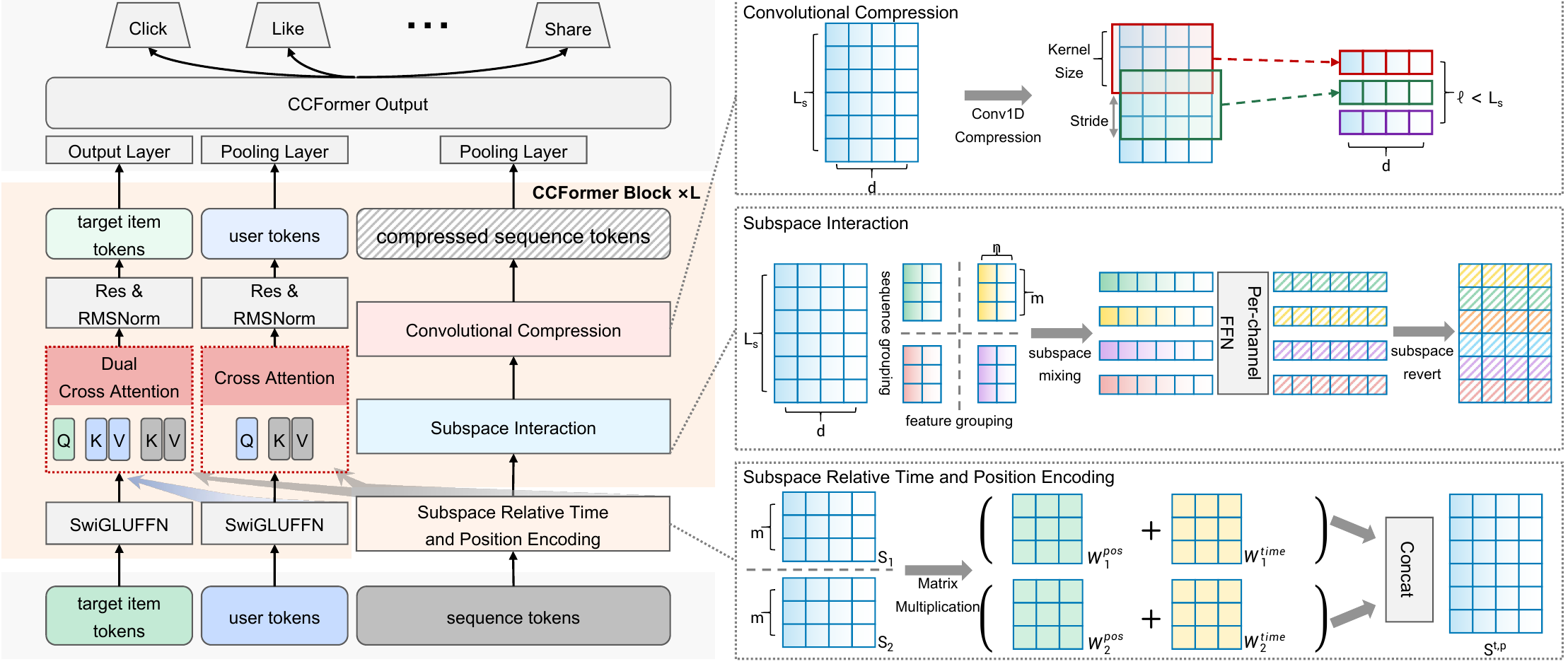}
\caption{Overview of CCFormer. The input feature space is partitioned into user, behavior-sequence, and target-item fields; each block combines feature-field separated cross attention, subspace token mixing with relative temporal-positional encoding, and convolutional sequence compression, so that global self-attention over the concatenated token set is never computed.}
\label{fig:ccformer_overview}
\end{figure}

CCFormer partitions the input feature space into three semantic fields---user-profile tokens $\mathbf{U} \in \mathbb{R}^{L_u \times d}$, behavior-sequence tokens $\mathbf{S} \in \mathbb{R}^{L_s \times d}$ (one fine-grained token per historical item), and target-item tokens $\mathbf{T} \in \mathbb{R}^{L_t \times d}$---and stacks $L$ interaction blocks that iteratively refine all three (Figure~\ref{fig:ccformer_overview}). After the final block, the normalized field representations are aggregated and fed to task-specific heads for multi-task prediction. Four mechanisms, applied per block, replace global self-attention:

\begin{enumerate}[leftmargin=1.4em,itemsep=2pt,topsep=2pt]
\item \textbf{Feature-field separated cross attention.} Instead of undirected attention over all concatenated tokens (which treats semantically different fields uniformly and wastes computation), three \emph{directed} attention flows implement the heterogeneous interactions of ranking: a user$\rightarrow$sequence flow (preference retrieval over the full history), a target$\rightarrow$sequence flow (target--history relevance matching), and a target$\rightarrow$user flow (target--user compatibility). RMSNorm aligns the scales of the fields and lightweight SwiGLU FFNs refine user and target tokens before crossing. Crucially, target tokens act only as queries and never attend to one another---a design choice that later enables single-pass scoring of all candidates (\S\ref{sec:genrank_infra}).
\item \textbf{Relative temporal-positional encoding in token subspaces.} Recency-aware order information is injected \emph{within local groups of $m$ behaviors}: a learnable time-decay weight $\alpha\,\beta^{|t_i - t_j|^{\gamma}}$ (with $\beta \in (0,1)$, so temporally proximate behaviors receive larger weights) plus a learnable relative positional bias, at $\mathcal{O}(L_s m)$ cost rather than the $\mathcal{O}(L_s^2)$ of attention-matrix biases.
\item \textbf{Long-sequence subspace token mixing.} The sequence tensor is reshaped into compact subspaces, each jointly spanning $m$ adjacent behavior tokens and $n$ hidden channels, and each channel group is processed by an independent gated per-channel feed-forward network. This forces token-level and channel-level signals to interact directly within localized subspaces, at cost linear in $L_s$. To our knowledge this is the first attempt to extend the token-mixing paradigm of industrial ranking \citep{rankmixer_2025} from small sets of unified feature tokens to full-length behavior sequences.
\item \textbf{Hierarchical sequence compression.} After each block, a strided one-dimensional convolution (kernel $k$, stride $s$) fuses adjacent behavior tokens and hands a shortened sequence to the next block. The receptive field of each surviving token thus expands progressively with depth: shallow layers capture short-term, fine-grained patterns over the raw sequence, while deep layers extract abstract, long-term preference signals over the full history. Because later blocks operate on geometrically shorter sequences, total computation drops substantially while access to the entire sequence is preserved---compression \emph{across} layers complements mixing \emph{within} layers.
\end{enumerate}

\subsection{Training and Serving at Industrial Scale}
\label{sec:genrank_infra}

CCFormer's deployment recipe is as much a part of TGR-GenRank as its architecture; four ingredients carry the model from research code to main-traffic serving.

\paragraph{Numerous-Torch infrastructure.} CCFormer is built on top of \emph{Numerous-Torch}, Tencent's production-grade training and inference infrastructure for large-scale recommendation models, whose role is to bridge flexible model innovation and industrial-scale deployment. On the modeling side it keeps a PyTorch-native development experience, so emerging advances from the rapidly evolving LLM community can be reused directly; on the system side it supplies distributed data ingestion, large-scale sparse-feature processing, sparse--dense joint training, checkpoint recovery, model export, and continual training on fresh user-feedback logs.

\paragraph{Mixed-precision training.} The main model computation runs in BF16/FP16 mixed precision, reducing memory overhead by more than $35\%$---headroom that is spent on larger batches or larger models under the same training budget.

\paragraph{Sparse-parameter compression.} Sparse user- and item-ID embeddings dominate the parameter count of industrial rankers. CCFormer stores ID-feature embedding tables with INT8 symmetric linear quantization ($\sim$$70\%$ compression over full-precision storage) and shrinks the tables themselves with a double-hashing strategy (a further $\sim$$50\%$ reduction in sparse parameter size). Together these substantially cut memory footprint and communication cost during training without degrading training effectiveness.

\paragraph{Single-pass parallel scoring of candidates.} Because target tokens never attend to each other, there is no cross-target information leakage, and all candidate items of a request can be packed into the target field and forwarded \emph{once}: user-field and sequence-field computation is shared across candidates, and the model emits multi-task scores for every candidate in a single pass. Relative to the pointwise per-candidate inference of the incumbent DLRM, this raises online peak QPS by $30\%$ at unchanged inference resources---even though CCFormer's per-sample computational cost is $20\times$ higher.

\paragraph{Default configuration.} Industrial experiments use feature-token dimensionality $d=256$, $8$ CCFormer blocks, subspace group sizes $m=8$ and $n=16$, compression kernel $k=3$ with stride $s=2$, Adam with learning rate $10^{-4}$, and batch size $4096$; all offline comparisons below run on an identical hardware/software stack of 16 NVIDIA H20 GPUs. Per-model default configurations across the stack are consolidated in Appendix~\ref{app:configs}.

\subsection{Offline Experiments}
\label{sec:genrank_offline}

\paragraph{Setup.} We evaluate on two public benchmarks---Taobao\footnote{\url{https://tianchi.aliyun.com/dataset/649}} and KuaiRec \citep{kuairec_2022}, with behavior sequences truncated at length $200$---and an industrial dataset of one month of production logs from a Tencent recommendation system: over $4$ billion samples, $30$ million users, and $10$ million items, with behavior sequence length $1000$. Public-benchmark baselines span the conventional DLRM paradigm (DIN \citep{din_2018}, DeepFM \citep{deepfm_2017}, SASRec \citep{sasrec_2018}, MIMN \citep{mimn_2019}) and attention-based sequential rankers (HSTU \citep{hstu_2024}, OneTrans \citep{onetrans_2025}, STCA \citep{stca_2026}); the industrial comparison uses the production-oriented subset (HSTU, OneTrans, STCA) with both AUC and GAUC. $\Delta$AUC / $\Delta$GAUC denote relative improvements in the RelaImpr sense, and every method receives its own TPE-based hyperparameter search for fairness.

\begin{table}[t]
\centering
\caption{Offline comparison on public benchmarks (mean $\pm$ std over 5 runs). $\Delta$AUC is the relative improvement over the DIN base model. Best in \textbf{bold}, runner-up \underline{underlined}.}
\label{tab:ccformer_public}
\small
\begin{tabular}{l|cc|cc}
\toprule
& \multicolumn{2}{c|}{\textbf{Taobao}} & \multicolumn{2}{c}{\textbf{KuaiRec}} \\
Method & AUC (\%) & $\Delta$AUC (\%) & AUC (\%) & $\Delta$AUC (\%) \\
\midrule
DIN      & 88.33$\pm$0.22 & 0.00  & 80.22$\pm$0.58 & 0.00 \\
DeepFM   & 89.06$\pm$0.42 & 1.90  & 80.14$\pm$0.51 & $-$0.26 \\
SASRec   & 88.52$\pm$0.65 & 0.50  & 79.83$\pm$0.70 & $-$1.29 \\
MIMN     & 91.79$\pm$0.32 & 9.03  & 81.79$\pm$0.81 & 5.20 \\
HSTU     & 91.40$\pm$0.25 & 8.01  & 82.62$\pm$0.39 & 7.94 \\
OneTrans & 91.35$\pm$0.61 & 7.88  & \underline{82.76$\pm$0.59} & \underline{8.41} \\
STCA     & \underline{92.81$\pm$0.50} & \underline{11.69} & 82.18$\pm$0.16 & 6.49 \\
CCFormer & \textbf{93.67$\pm$0.32} & \textbf{13.93} & \textbf{83.35$\pm$0.29} & \textbf{10.36} \\
\bottomrule
\end{tabular}
\end{table}

\begin{table}[t]
\centering
\caption{Offline comparison on the industrial dataset ($4$B+ samples, sequence length $1000$). $\Delta$ columns are relative improvements over the HSTU base model. Best in \textbf{bold}, runner-up \underline{underlined}.}
\label{tab:ccformer_industrial}
\small
\begin{tabular}{l|cc|cc}
\toprule
Model & AUC (\%) & $\Delta$AUC (\%) & GAUC (\%) & $\Delta$GAUC (\%) \\
\midrule
HSTU     & 77.66 & 0.00 & 70.86 & 0.00 \\
OneTrans & 77.69 & 0.11 & 70.90 & 0.19 \\
STCA     & \underline{77.73} & \underline{0.25} & \underline{70.95} & \underline{0.43} \\
CCFormer & \textbf{77.94} & \textbf{1.01} & \textbf{71.36} & \textbf{2.40} \\
\bottomrule
\end{tabular}
\end{table}

\paragraph{Overall results.} CCFormer achieves the best score on every dataset and metric (Tables~\ref{tab:ccformer_public} and~\ref{tab:ccformer_industrial}). On Taobao and KuaiRec it reaches $93.67\%$ and $83.35\%$ AUC, beating the strongest sequential baseline on each dataset (STCA and OneTrans, respectively) by $0.86$ and $0.59$ points, and it maintains a consistent margin over HSTU, OneTrans, and STCA when the KuaiRec sequence length is varied over $\{200, 500, 1000\}$. On the industrial dataset it improves over HSTU by $+0.28$ AUC and $+0.50$ GAUC points ($1.01\%$ and $2.40\%$ relative)---a large margin at production scale, where $0.1\%$-level AUC moves are considered shippable---and still clears the strongest baseline STCA by $+0.21$ AUC and $+0.41$ GAUC points.

\paragraph{Hyperparameter robustness.} The two long-sequence-specific hyperparameter families are insensitive over wide ranges: varying the compression kernel $k$ from $2$ to $7$ moves AUC only within $77.92$--$77.95\%$, and all tested $(m, n)$ subspace group sizes outperform HSTU with limited variation. In practice, CCFormer transfers to new scenarios without delicate tuning.

\subsection{Scaling Behavior}
\label{sec:genrank_scaling}

The premise of generative-paradigm ranking is that ranking quality can be bought by FLOPs; the question is at what exchange rate. We examine the two axes that matter most for sequential rankers---sequence length and model capacity---holding the hardware/software stack fixed and reporting GFLOPs alongside quality.

\begin{figure}[t]
\centering
\includegraphics[width=0.78\textwidth]{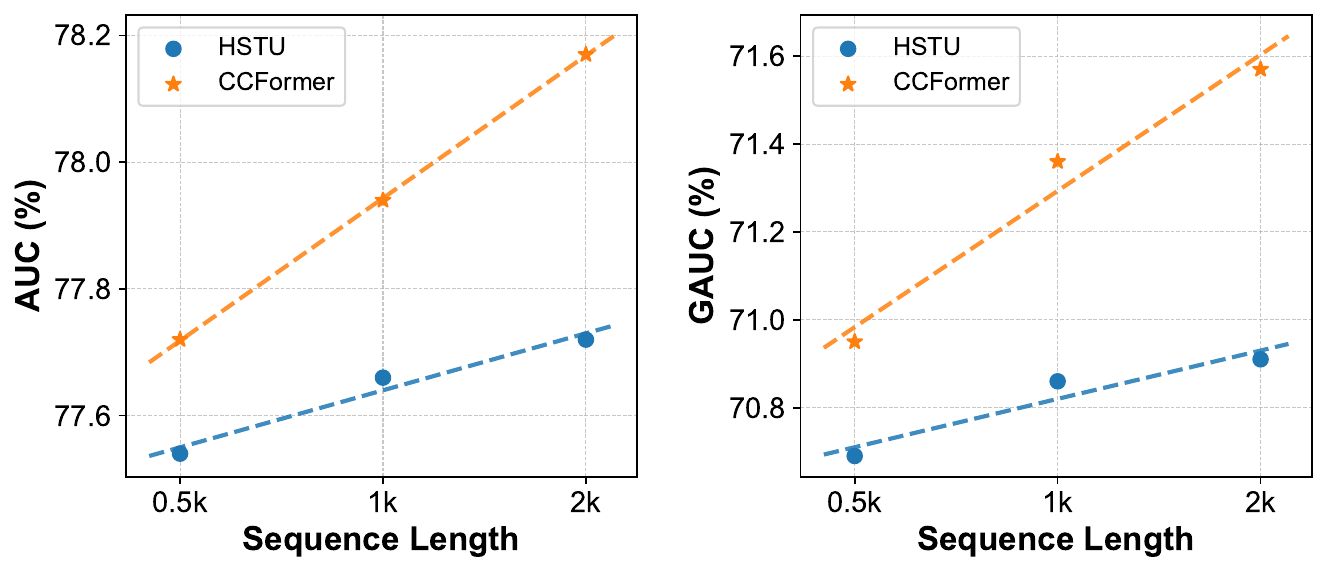}
\caption{Sequence-length scaling on the industrial dataset (lengths $500$, $1000$, $2000$). CCFormer improves predictably with longer sequences and dominates HSTU at every length; with $0.5$k behavior tokens it already matches the AUC and surpasses the GAUC of HSTU consuming $2$k tokens.}
\label{fig:ccformer_scaling}
\end{figure}

\paragraph{Sequence-length scaling.} As the industrial sequence length grows from $0.5$k to $2$k, CCFormer improves monotonically from $77.72\%$ to $78.17\%$ AUC and from $70.95\%$ to $71.57\%$ GAUC, with average relative gains over HSTU of $1.08\%$ AUC and $2.37\%$ GAUC across length settings---and the margin \emph{widens} as sequences lengthen (Figure~\ref{fig:ccformer_scaling}). Most strikingly, \textbf{CCFormer with $0.5$k behavior tokens already matches the AUC and surpasses the GAUC of HSTU with $2$k tokens}: subspace token mixing extracts expressive patterns from the full history, while hierarchical compression preserves a large receptive field at a fraction of the cost.

\paragraph{Model-size scaling.} Varying the feature dimensionality over $\{128, 256, 512\}$ (Table~\ref{tab:ccformer_size}), CCFormer again scales predictably ($77.65\% \to 78.13\%$ AUC, $70.76\% \to 71.52\%$ GAUC, average relative gains of $1.02\%$ AUC and $1.93\%$ GAUC over HSTU at matched capacity) while consuming roughly half the GFLOPs of HSTU at every capacity point---at $d=128$ it comes within $0.01$ AUC points of HSTU at $d=256$ using $4\times$ fewer GFLOPs per sample. The two analyses together show that CCFormer offers a strictly better effectiveness--efficiency exchange rate, which is what makes scaling \emph{deployable} rather than merely demonstrable.

\begin{table}[t]
\centering
\caption{Model-size scaling on the industrial dataset. $\Delta$ columns are relative improvements over HSTU at the same feature dimensionality.}
\label{tab:ccformer_size}
\small
\begin{tabular}{l|c|cc|cc|c}
\toprule
Model & Dim. & AUC (\%) & $\Delta$AUC (\%) & GAUC (\%) & $\Delta$GAUC (\%) & GFLOPs/sample \\
\midrule
HSTU     & 128 & 77.42 & 0.00 & 70.55 & 0.00 & 28.87 \\
CCFormer & 128 & 77.65 & 0.84 & 70.76 & 1.02 & \phantom{0}9.34 \\
\midrule
HSTU     & 256 & 77.66 & 0.00 & 70.86 & 0.00 & 38.37 \\
CCFormer & 256 & 77.94 & 1.01 & 71.36 & 2.40 & 18.28 \\
\midrule
HSTU     & 512 & 77.79 & 0.00 & 71.02 & 0.00 & 63.47 \\
CCFormer & 512 & 78.13 & 1.22 & 71.52 & 2.38 & 36.14 \\
\bottomrule
\end{tabular}
\end{table}

\paragraph{Ablations: where the speed and the quality come from.} Table~\ref{tab:ccformer_ablation} disentangles the contributions of the three core designs, measuring training speedup as relative training throughput (samples per unit time) against HSTU under an identical stack. Three findings stand out. First, \emph{sequence compression is the efficiency lever}: removing it leaves AUC essentially unchanged ($77.96\%$) but collapses the training speedup from $2.21\times$ to $1.29\times$. Second, \emph{subspace token mixing is the quality workhorse}: removing it causes the largest degradation ($-0.21$ AUC / $-0.34$ GAUC points), while replacing it with standard self-attention buys only $+0.02$ AUC points, \emph{loses} $0.15$ GAUC points, and drops the speedup to $1.41\times$---costly global attention over the sequence is unnecessary. Third, relative temporal-positional encoding contributes $+0.09$ AUC / $+0.23$ GAUC points by separating recent behaviors from outdated ones.

\begin{table}[t]
\centering
\caption{Ablation study on the industrial dataset. Training speedup is the training throughput relative to HSTU; $\Delta$ columns are relative improvements over HSTU. Best in \textbf{bold}, runner-up \underline{underlined}.}
\label{tab:ccformer_ablation}
\small
\setlength{\tabcolsep}{3.5pt}
\begin{tabular}{l|c|cc|cc}
\toprule
Variant & Speedup & AUC (\%) & $\Delta$AUC (\%) & GAUC (\%) & $\Delta$GAUC (\%) \\
\midrule
HSTU (base)                                & 1.00$\times$ & 77.66 & 0.00 & 70.86 & 0.00 \\
\midrule
CCFormer (full)                            & 2.21$\times$ & \underline{77.94} & \underline{1.01} & \textbf{71.36} & \textbf{2.40} \\
\quad w/o relative temporal-pos.\ encoding & \underline{2.35}$\times$ & 77.85 & 0.69 & 71.13 & 1.29 \\
\quad token mixing $\rightarrow$ self-attn & 1.41$\times$ & \textbf{77.96} & \textbf{1.08} & 71.21 & 1.68 \\
\quad w/o long-seq.\ token mixing          & \textbf{2.40}$\times$ & 77.73 & 0.25 & 71.02 & 0.77 \\
\quad w/o sequence compression             & 1.29$\times$ & \textbf{77.96} & \textbf{1.08} & \underline{71.25} & \underline{1.87} \\
\bottomrule
\end{tabular}
\end{table}

\subsection{Online Deployment and A/B Results}
\label{sec:genrank_online}

\begin{table*}[!htbp]
    \centering
    \setlength{\tabcolsep}{4.5pt}
    \caption{Online A/B gains of CCFormer across five Tencent production scenarios. Each experimental group served $>$1M exposed users per day over a two-week window; all lifts are statistically significant (two-sample $t$-test, $p<0.05$).}\label{tab:ab_test}
    \begin{tabular}{c|cccccccc}
        \toprule
        \multicolumn{9}{c}{\textbf{Scenario 1}} \\
        \midrule
        \textbf{Metric} & CTR & UCTR & \makecell{3-day\\Activeness} & \makecell{Video\\View} & \makecell{Unique\\Viewer} & \makecell{Page\\View} & \makecell{Watch\\Time} & \makecell{Ad\\revenue} \\ 
        \cmidrule{2-9}
        \textbf{Lift (\%)} & 3.57 & 1.93 & 1.93 & 3.47 & 2.61 & 3.86 & 1.29 & 1.64 \\
        \midrule
        \multicolumn{9}{c}{\textbf{Scenario 2}} \\
        \midrule
        \textbf{Metric} & \multicolumn{4}{c}{Ad view} & \multicolumn{4}{c}{Ad revenue} \\
        \cmidrule{2-9}
        \textbf{Lift (\%)} & \multicolumn{4}{c}{1.43} & \multicolumn{4}{c}{1.71} \\
        \midrule
        \multicolumn{9}{c}{\textbf{Scenario 3}} \\
        \midrule
        \textbf{Metric} & \multicolumn{2}{c}{CTR} & \multicolumn{2}{c}{UCTR} & \multicolumn{2}{c}{New User Page View} & \multicolumn{2}{c}{Ad revenue}  \\
        \cmidrule{2-9}
        \textbf{Lift (\%)} & \multicolumn{2}{c}{2.02} & \multicolumn{2}{c}{1.67} & \multicolumn{2}{c}{4.01}  & \multicolumn{2}{c}{1.29} \\
        \midrule
        \multicolumn{9}{c}{\textbf{Scenario 4}} \\
        \midrule
        \textbf{Metric} & \multicolumn{2}{c}{Page View} & \multicolumn{2}{c}{Finish} & \multicolumn{2}{c}{Watch Time} & \multicolumn{2}{c}{Follow} \\
        \cmidrule{2-9}
        \textbf{Lift (\%)} & \multicolumn{2}{c}{0.55} & \multicolumn{2}{c}{0.98} & \multicolumn{2}{c}{0.42} & \multicolumn{2}{c}{2.18} \\
        \midrule
        \multicolumn{9}{c}{\textbf{Scenario 5}} \\
        \midrule
        \textbf{Metric} & \multicolumn{2}{c}{Page View} & \multicolumn{2}{c}{Watch Time} & \multicolumn{2}{c}{Share} & \multicolumn{2}{c}{CTR} \\
        \cmidrule{2-9}
        \textbf{Lift (\%)} & \multicolumn{2}{c}{0.45} & \multicolumn{2}{c}{1.26} & \multicolumn{2}{c}{3.49} & \multicolumn{2}{c}{1.01}\\
        \bottomrule
    \end{tabular}
\end{table*}

CCFormer has been A/B-tested in five Tencent scenarios covering video recommendation, content feeds, and advertising ranking. The incumbents differ by scenario, which makes the comparison informative in two directions: Scenario~1 replaces a long-standing, iteratively optimized DLRM ranker, whereas Scenarios~2--5 replace an already deployed HSTU model. Each experimental group served more than one million exposed users per day over a two-week window. Every lift reported is statistically significant (two-sample $t$-test, $p<0.05$), confirming the reliability of the improvements in the production environment.

In Scenario~1 (video recommendation), where CCFormer replaces the long-standing production DLRM, the gains span the full engagement funnel---click (CTR $+3.57\%$, UCTR $+1.93\%$), consumption (Page View $+3.86\%$, Video View $+3.47\%$, Watch Time $+1.29\%$), audience (Unique Viewer $+2.61\%$), and short-term retention (3-day Activeness $+1.93\%$)---and both Scenario~1 and the advertising-ranking Scenario~2 lift advertising revenue ($+1.64\%$ and $+1.71\%$), the key monetization metric. Following the A/B tests, CCFormer was fully launched in Scenarios~1 and~2 and now serves the entire production traffic of both, with the gains remaining stable after full deployment.

\section{TGR-GenRec: From Next-Token to Next-Slate Generation}
\label{sec:higr}

This section presents \textbf{TGR-GenRec}, the end-to-end generation direction of the TGR Stack and its most fully developed component. TGR-GenRec is a \emph{direction}, not a single model: it spans the two generation paradigms of generative recommendation, and currently ships one production model for each---\textbf{BARGE} \citep{barge_2027} under the \emph{next-token-prediction} (NTP) paradigm that generates one item at a time, and \textbf{HiGR} \citep{higr_2026} under the \emph{slate-generation} paradigm that emits the whole result list as one structured object. We first position the two paradigms and models (\S\ref{sec:genrec_overview}); \S\ref{sec:barge} and \S\ref{sec:higr_model} then present the two works independently, as parallel and self-contained units, each with its own motivation, method, offline results, and deployment.

\subsection{Two Generation Paradigms: Next-Token and Next-Slate Prediction}
\label{sec:genrec_overview}

The long-term goal of TGR-GenRec is to replace the cascade's learned decision stages with generative models that directly produce the ordered recommendation result. The deployments reported here are intermediate realizations of that goal rather than a uniform full-pipeline replacement. In production this goal is pursued through two generation paradigms that pose logically independent problems, attacked by two separate works. We present them in order of generation granularity---from single-item NTP to whole-slate generation:

\begin{itemize}[leftmargin=1.4em,itemsep=2pt,topsep=2pt]
\item \textbf{The NTP paradigm --- generating one item faithfully (BARGE, \S\ref{sec:barge}).} The mainstream generative formulation (the TIGER lineage) tokenizes each item into a hierarchical SID tuple and predicts the next item token by token. But this formulation was borrowed from natural language, where every token carries stand-alone semantics; a hierarchical SID codeword is meaningful only jointly with its prefix. The mismatch creates two structural gaps in \emph{any} NTP-style SID generator: flattening dissolves item boundaries at the encoder, and hierarchical decoding accumulates semantic drift---an error at any layer strands the search in a wrong subtree. BARGE's answer is to adapt the NTP mechanism itself to the recommendation task: restore item structure at the encoder (ICA), rerank decoding paths for global coherence (HPR), and decode through two orthogonal quantization channels (DPD).
\item \textbf{The slate-generation (next-slate-prediction, NSP) paradigm --- generating the whole list (HiGR, \S\ref{sec:higr_model}).} Even a perfectly faithful next-item generator decides items one at a time, while our surfaces present an ordered slate in a single display: quality lives at the list level---ordering, compatibility, diversity---and decoding $M \times D$ tokens item by item is too slow under real-time constraints, with token-level likelihood optimizing neither. HiGR's answer is to make the \emph{list} the unit of generation: hierarchical semantic IDs with controllable prefixes (PCRQ-VAE), coarse-to-fine slate planning and item decoding (HSD), and listwise multi-objective alignment (ORPO).
\end{itemize}

The two models are orthogonal by construction---BARGE hardens the \emph{generation mechanism} within the NTP paradigm, HiGR changes the \emph{output structure and objectives} of generation---and their fixes act on disjoint failure modes. Table~\ref{tab:genrec_models} summarizes the positioning; both were shipped under the same production discipline (principle~P1, \S\ref{sec:overview_principles}): drop-in deployment behind the existing filter stack at matched or better latency and GPU budgets.

\begin{table}[t]
\centering
\caption{The two TGR-GenRec production models, one per generation paradigm. BARGE and HiGR arise from different motivations and repair complementary failure modes of end-to-end generative recommendation.}
\label{tab:genrec_models}
\small
\setlength{\tabcolsep}{5pt}
\resizebox{\textwidth}{!}{%
\begin{tabular}{l|l|l}
\toprule
                        & \textbf{BARGE} \citep{barge_2027} & \textbf{HiGR} \citep{higr_2026} \\
\midrule
Generation paradigm     & Next-token prediction (single item)      & Slate generation (whole list) \\
Motivating gap          & NLP-born autoregression structurally     & Item-level generation misaligns \\
                        & mismatches SID-based recommendation      & with list-level UX and latency \\
Granularity attacked    & Token/path level (\emph{how} to generate) & Slate level (\emph{what} to generate) \\
Failure modes repaired  & Item-boundary loss; hierarchical    & Entangled SID prefixes; $M{\times}D$-token \\
                        & semantic drift                      & decoding cost; tokenwise objectives \\
Tokenizer               & OSQ-VAE (orthogonal dual-channel)        & PCRQ-VAE (prefix-contrastive) \\
Decoder                 & Dual-decoder + HPR reranking, OR-fusion  & HSD: slate planner + item generator \\
Training / alignment    & NTP + symmetric InfoNCE (label-free)     & NTP + ORPO listwise (3 objectives) \\
Serving-cost delta      & Zero (params, beam, latency unchanged)   & $-60\%$ GPU; P99 $<50$\,ms \\
Online A/B              & $+0.60\%$ CTR, $+1.34\%$ unique clicking users,       & $+1.22\%$ watch time, $+1.73\%$ views \\
                        & $+1.70\%$ reading time (6\% traffic)     & (5\% traffic) \\
\bottomrule
\end{tabular}}
\end{table}

\subsection{BARGE: Adapting Next-Token Prediction for Industrial Recommendation}
\label{sec:barge}

\textbf{BARGE} (\textbf{B}ridging \textbf{A}uto\textbf{R}egressive \textbf{G}eneration for r\textbf{E}commendation) \citep{barge_2027} is the TGR-GenRec production model under the \emph{next-token-prediction} paradigm (\S\ref{sec:genrec_overview}): each item is represented as a hierarchical SID tuple and the next item is generated token by token, in the mainstream TIGER-style formulation. Rather than replacing this paradigm, BARGE hardens \emph{how} SID sequences are generated, closing two structural gaps between the NLP-born autoregressive formulation and the recommendation task. BARGE was designed under a hard production constraint---\emph{no increase in parameters, beam budget, or serving latency}---and has been deployed on Tencent's commercial media platform, where a 6\% live-traffic A/B test delivered $+0.60\%$ click-through rate, $+1.34\%$ unique clicking users, and $+1.70\%$ total reading time over the incumbent multi-stage system.

\subsubsection{Motivation: Two Structural Gaps at Industrial Scale}
\label{sec:barge_motivation}

Generative recommendation avoids exhaustive scoring over the entire item corpus by representing each item as a sequence of hierarchical semantic IDs (SIDs) and autoregressively generating the target sequence. Although this formulation enables efficient token-level generation, it introduces a mismatch between the modeling unit and the task unit: the backbone operates on individual SID tokens, whereas user interactions and recommendation targets are defined at the item level. In RQ-VAE-based implementations, these SIDs form an ordered, coarse-to-fine code path. Flattening these paths removes explicit item grouping from the encoder, while autoregressive decoding exposes the hierarchical path to cumulative prefix errors. These properties give rise to two structural gaps in the encode--decode pipeline, which become particularly consequential at industrial scale.

\begin{figure*}[t]
    \centering
    \includegraphics[width=0.98\textwidth]{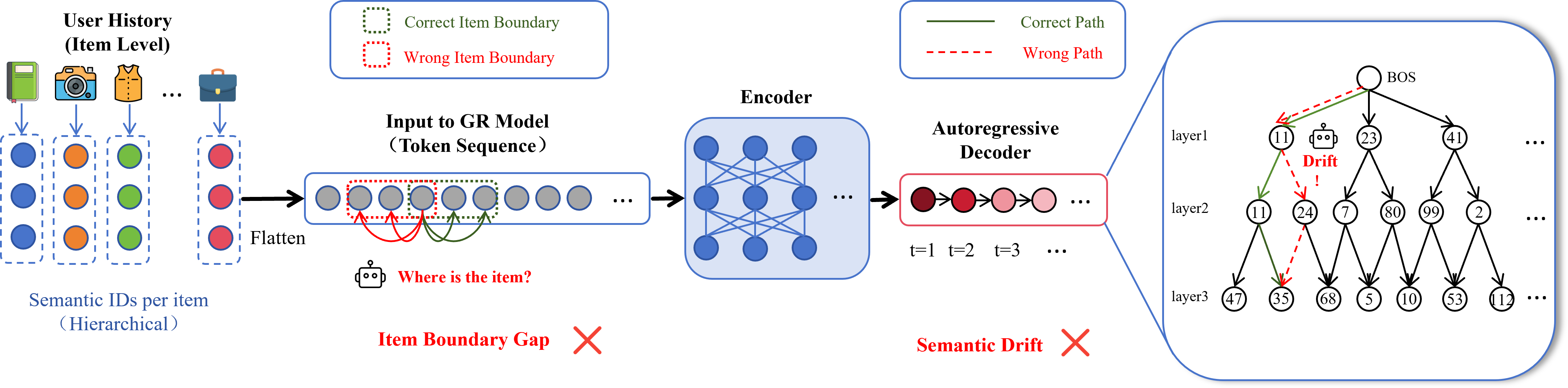}
    \caption{
    Two structural gaps in generative recommendation.
    \textbf{Left (Item-Boundary Gap, P1):}
    after each item is tokenized into $L$ hierarchical semantic IDs
    and flattened into a single sequence, item boundaries vanish from
    the input to the GR model, and the encoder is no longer explicitly 
    informed which tokens belong to the same item (``Where is the item?'').
    \textbf{Right (Semantic Drift, P2):}
    on the hierarchical codebook tree, an error at any layer redirects
    decoding into a wrong subtree, so that the target leaf becomes
    unreachable along the selected path no matter how the remaining
    tokens are scored.
    }
    \label{fig:barge_structural_gaps}
\end{figure*}

\paragraph{(P1) Item-boundary gap.} Existing RQ-VAE-based generative recommenders \citep{tiger_2023,letter_2024} tokenize each item into an $L$-level SID and concatenate all SIDs in the user history into a single token sequence. Although this representation is convenient for standard sequence models, it removes explicit item boundaries before the sequence enters the encoder. Self-attention therefore processes all SID tokens in the same manner and must infer which tokens belong to the same item mainly from positional and contextual signals.

A more subtle ambiguity arises from the hierarchical nature of RQ-VAE codewords. The semantics of a codeword depend on its prefix, whereas its input embedding is typically determined only by its code index and quantization level. Consider two items with SIDs $\langle 1,3,5\rangle$ and $\langle 7,3,8\rangle$. At the second level, both items use codeword $3$ and therefore receive the same initial lookup embedding for that token. However, the partial paths $\langle 1,3\rangle$ and $\langle 7,3\rangle$ correspond to different nodes in the hierarchical codebook and need not represent the same item-level semantics. The downstream encoder must consequently recover this prefix-conditioned distinction through contextualization.

After flattening, the encoder must simultaneously reconstruct item membership and disambiguate the prefix-dependent semantics of individual codewords. This places the burden of item-level preference modeling on implicit token interactions, even though recommendation is fundamentally performed at the item level. As illustrated on the left of Figure~\ref{fig:barge_structural_gaps}, this mismatch motivates an explicit item-level aggregation mechanism that preserves token-level information while restoring the structural context shared by the SID tokens of each item.

\paragraph{(P2) Semantic drift.} Hierarchical SIDs define a tree-structured decoding path, in which each predicted codeword constrains the subtrees reachable at subsequent layers. An early error therefore redirects decoding to an incorrect subtree and makes the target item unreachable along the selected path. Although beam search retains multiple partial paths, it ranks them mainly by accumulated token log-probabilities without explicitly evaluating their global semantic coherence. We refer to this prefix-induced error propagation as \emph{semantic drift}, as illustrated on the right of Figure~\ref{fig:barge_structural_gaps}.

Table~\ref{tab:barge_drift} quantifies this effect on the TIGER baseline. At the deepest layer $c_3$, the probability assigned to the target codeword decreases from $0.787$ under teacher forcing to $0.015$ under autoregressive decoding, a roughly $52\times$ drop. Under autoregressive decoding, the $c_3$ accuracy is $77.0\%$ with a correct prefix but only $0.6\%$ with an erroneous one, yielding a gap of more than $128\times$. This result shows that deep-layer failures arise largely from preceding path errors, rather than solely from the prediction difficulty of the current layer.

Crucially for production, the brute-force remedy---enlarging the beam width $B$---is not viable: inference latency scales approximately linearly with $B$, while beam-dependent decoding memory grows with $B\times L$. Moreover, a larger beam only preserves more candidate paths without explicitly correcting the underlying semantic drift. This trade-off is particularly acute at platform scale, where the item corpus can contain tens or even hundreds of millions of items and follows a highly skewed long-tail distribution, making path-consistent decoding more important under strict online serving constraints.

\begin{table}[t]
\centering
\caption{Per-layer prediction quality under Teacher Forcing (TF) and Autoregressive (AR) decoding on Amazon Beauty (TIGER baseline, greedy decoding). The last two rows split AR results at $c_3$ by prefix correctness, exposing the error cascade that motivates BARGE.}
\label{tab:barge_drift}
\small
\begin{tabular}{l|ccc}
\toprule
Layer / Mode & Mismatch $\downarrow$ & Rank $\downarrow$ & Prob $\uparrow$ \\
\midrule
$c_1$ (TF / AR)            & 0.915 & 66.9  & 0.046 \\
$c_2$ (TF)                 & 0.864 & 25.3  & 0.102 \\
$c_2$ (AR)                 & 0.981 & 108.2 & 0.015 \\
$c_3$ (TF)                 & 0.195 & 7.4   & 0.787 \\
$c_3$ (AR)                 & 0.984 & 119.5 & 0.015 \\
\midrule
$c_3$ (AR, prefix correct) & 0.230 & 3.7   & 0.738 \\
$c_3$ (AR, prefix error)   & 0.994 & 121.1 & 0.006 \\
\bottomrule
\end{tabular}
\end{table}

Together, these observations motivate BARGE. Item Context-Aware Attention (ICA) restores item-level structure in the encoder, Hierarchical Path Reranking (HPR) evaluates global path coherence within each decoding channel, and Dual-Path Decoding (DPD) provides a structurally complementary recovery path. BARGE thereby addresses the two structural gaps without requiring a larger model or a wider beam.

\subsubsection{Method Overview}
\label{sec:barge_method}

\begin{figure}[t]
\centering
\includegraphics[width=0.98\textwidth]{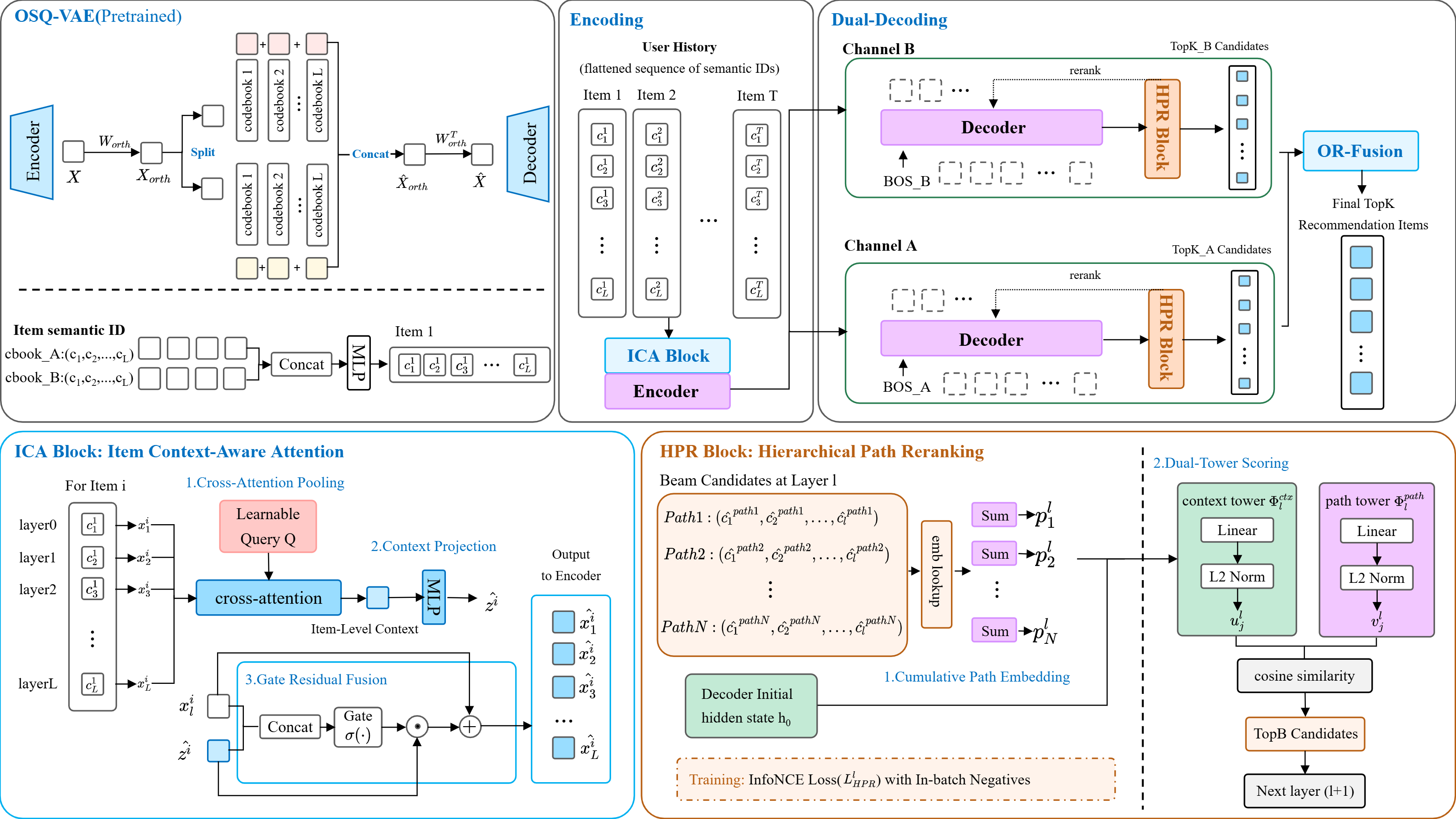}
\caption{Overview of BARGE. The OSQ-VAE tokenizes each item into two channel-specific SID tuples via orthogonal rotation and dual codebook stacks; the user history passes through the ICA block before the shared Transformer encoder; two decoders, each followed by an HPR block, run in parallel and their Top-$K$ lists are merged by OR-fusion.}
\label{fig:barge_overview}
\end{figure}

As shown in Figure~\ref{fig:barge_overview}, BARGE addresses the two structural gaps through three lightweight and complementary modules:

\begin{itemize}[leftmargin=1.4em,itemsep=3pt,topsep=3pt]

\item \textbf{Item Context-Aware Attention (ICA)} addresses the item-boundary gap on the encoder side. For each item, ICA uses a learnable query to aggregate its $L$ SID-token embeddings through cross-attention, and injects the resulting item-level context into every token through a gated residual connection. When the gate approaches zero, the module recovers an identity path and preserves the original token representation. Empirically, the average gate activations across the four SID levels concentrate around $0.35$--$0.38$ on the two public datasets, indicating that ICA learns a moderate level of context injection rather than overwriting the token-level signal.

\item \textbf{Hierarchical Path Reranking (HPR)} addresses semantic drift within each decoding channel. At layer $l$, a dual-tower scorer evaluates the compatibility between the decoder's pre-generation hidden state $\mathbf{h}_0$ and the cumulative path embedding $\mathbf{p}^{(l)}=\sum_{j=1}^{l}\mathbf{e}_{c_j}$. HPR is jointly trained with the main model using symmetric InfoNCE: each positive pair matches $\mathbf{h}_0$ with its ground-truth cumulative path, while the negatives combine in-batch paths, high-probability non-ground-truth prefixes sampled from the next-token prediction distribution, and optional business-feedback negatives such as impressed-but-unclicked items. These prefix-aware negatives expose HPR to plausible drift patterns during training. At inference, HPR combines its path-compatibility score with the generation log-probability at weight $\lambda$ to rerank a Top-$N$ pool, while retaining only the Top-$B$ paths for the next layer. The outgoing beam width therefore remains unchanged, and the context-side representation is shared across all candidate paths to limit the additional scoring cost.

\item \textbf{Dual-Path Decoding (DPD)} addresses semantic drift from a complementary decoding direction. Its Orthogonal Split-and-Quantize VAE (OSQ-VAE) applies a learnable Householder-parameterized rotation $R$, with
$R^\top R=I$ by construction, and splits the rotated item representation into two orthogonal coordinate subspaces. The two subspaces are quantized by independent residual codebook stacks, producing two SID channels for each item. A shared encoder feeds two channel-specific decoder towers, each equipped with private output heads and HPR scorers. Their item-space scores are merged through LSE-based soft-OR fusion, allowing an item ranked highly by either channel to be recovered. DPD keeps the per-channel beam width and final Top-$K$ output length unchanged, although the two decoding branches introduce additional decoder computation relative to a single-channel model.

\end{itemize}

\begin{figure}[ht]
\centering
\begin{minipage}[b]{0.48\linewidth}
\centering
\includegraphics[width=\linewidth]{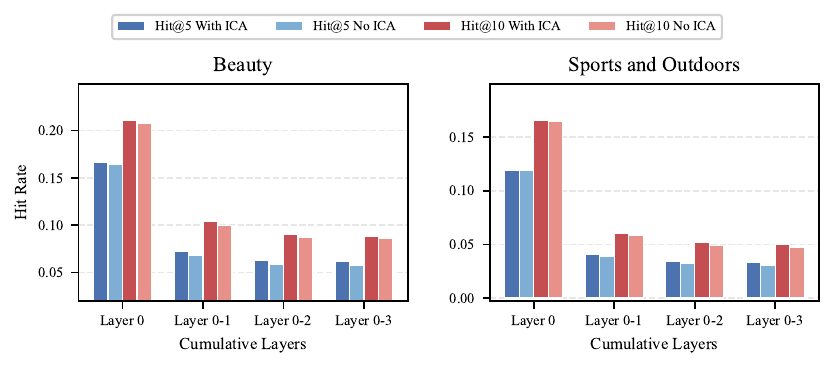}
\end{minipage}
\hfill
\begin{minipage}[b]{0.48\linewidth}
\centering
\includegraphics[width=\linewidth]{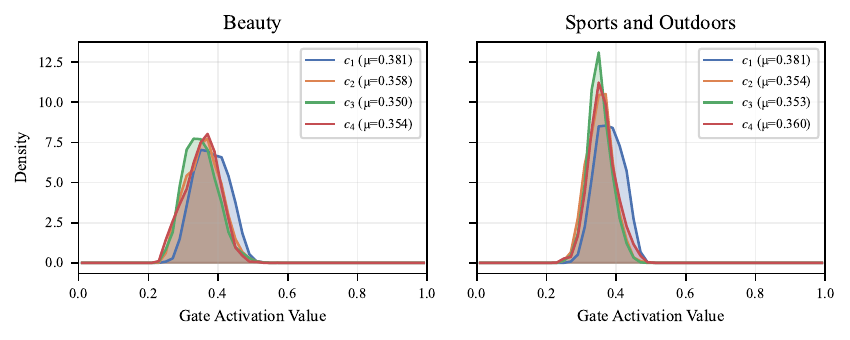}
\end{minipage}
\caption{\emph{Left:} cumulative hit rate with and without ICA across SID layers---the gap widens at deeper layers, showing stronger resistance to semantic drift. \emph{Right:} gate activation distribution across layers, stably concentrated around $0.35$--$0.38$, confirming selective and consistent injection of item-level context.}
\label{fig:barge_ica}
\end{figure}

The two decoding channels exhibit strong empirical complementarity. At $K=10$, their candidate pools have Jaccard similarities of only $0.183$ on Beauty and $0.172$ on Sports. Among the test cases successfully recalled by OR fusion, $15.6\%$ and $24.3\%$ are contributed exclusively by one channel on the two datasets, respectively. Figure~\ref{fig:barge_ica} further shows that the cumulative hit-rate advantage of ICA over its no-ICA counterpart increases at deeper SID levels, a trend consistent with improved resistance to hierarchical error propagation.

\subsubsection{Offline Results}
\label{sec:barge_offline}

\noindent\textbf{Public benchmark results.} On Amazon Beauty and Sports (leave-one-out, full-catalog evaluation), BARGE achieves the best score on every metric against both discriminative and generative baselines (Table~\ref{tab:barge_amazon}); e.g., on Beauty it lifts R@10 from $0.0775$ (ActionPiece) to $0.0927$ ($+19.6\%$) and beats TIGER by $43.0\%$. A codebook-controlled variant (BARGE-base) already surpasses all prior baselines, isolating the gains of the three structural modules from the codebook design. Component ablations show every module helps independently and their gains are largely additive, consistent with the modules acting on disjoint failure sources.

\begin{table}[t]
\centering
\caption{BARGE vs.\ the strongest published baselines on Amazon Beauty and Sports (R@10 / N@10). BARGE-base replaces the 4-layer learned codebook with TIGER's 3-layer codebook plus a collision-resolving ID.}
\label{tab:barge_amazon}
\small
\begin{tabular}{l|cc|cc}
\toprule
& \multicolumn{2}{c|}{\textbf{Beauty}} & \multicolumn{2}{c}{\textbf{Sports}} \\
Method & R@10 & N@10 & R@10 & N@10 \\
\midrule
TIGER \citep{tiger_2023}            & 0.0648 & 0.0384 & 0.0400 & 0.0225 \\
HSTU \citep{hstu_2024}              & 0.0704 & 0.0389 & 0.0414 & 0.0215 \\
COBRA \citep{cobra_2025}            & 0.0725 & 0.0456 & 0.0434 & 0.0257 \\
ActionPiece \citep{actionpiece_2025} & 0.0775 & 0.0424 & 0.0500 & 0.0264 \\
APAO-pointwise \citep{apao_2026}    & 0.0795 & 0.0453 & 0.0444 & 0.0237 \\
\midrule
BARGE-base                          & 0.0896 & 0.0515 & 0.0513 & 0.0285 \\
BARGE                               & \textbf{0.0927} & \textbf{0.0547} & \textbf{0.0544} & \textbf{0.0308} \\
\bottomrule
\end{tabular}
\end{table}

\noindent\textbf{Industrial-scale offline evaluation.} We evaluate BARGE in two large-scale recommendation scenarios on Tencent's commercial media platform. The two scenarios differ in both corpus scale and incumbent architecture, allowing us to examine BARGE under distinct production settings. Scenario 1 uses an $11$-day observation window, with the first $10$ days used for training and the final day for evaluation. The resulting dataset covers millions of users, hundreds of millions of interactions, and hundreds of thousands of items. Scenario 2 operates at a substantially larger scale, covering tens of millions of users, tens of millions of candidate items, and tens of billions of user--item interactions.

The deployed production baselines also differ by scenario. Scenario 1 compares BARGE with a GraphSAGE-based retrieval model and Neural Approximate Nearest Neighbor Search (NANN), while Scenario 2 uses a GraphSAGE-based retrieval model and a BERT-based sequential recommender. These models are production baselines currently deployed in their respective scenarios. In addition, we compare BARGE with OneRec as a generative baseline in both scenarios. Because the source code and model checkpoints corresponding to the production system described in the original OneRec technical report \citep{onerec_tr_2025} are not publicly available, we implement OneRec in-house following its published architecture and methodological description. This in-house OneRec baseline is used only for offline comparison and is not deployed in either scenario.

As shown in Table~\ref{tab:barge_industrial}, BARGE consistently achieves the best performance across all four Hit@$K$ metrics in both scenarios. In Scenario 1, BARGE improves Hit@5 by $10.2\%$ and Hit@10 by $6.2\%$ relative to the strongest baseline, OneRec. The gain is even more pronounced in Scenario 2, where BARGE improves Hit@5 by $16.9\%$ and Hit@10 by $14.7\%$ over OneRec. Its relative improvement remains $12.6\%$ at Hit@50, showing that the gain extends beyond the first few positions of the ranked list.

The compared systems span graph-based retrieval, Neural Approximate Nearest Neighbor Search, BERT-based sequential recommendation, and semantic-ID-based generative recommendation. GraphSAGE-based, NANN, and BERT-based models operate primarily on atomic item representations and do not explicitly model the coarse-grained semantic structure shared by related items. OneRec narrows the gap through semantic-ID-based generation, but its single-path token-level objective remains vulnerable to shallow-layer errors and off-path items. By combining item-level structure modeling, path-level reranking, and complementary decoding channels, BARGE delivers consistent improvements across both industrial scenarios. These results suggest that its structural adaptations generalize across different corpus scales and incumbent architectures.

\begin{table}[t]
\centering
\caption{Offline performance in two large-scale recommendation scenarios
on Tencent's commercial media platform. Underlined and bold values denote
the best baseline and the best overall result, respectively.}
\label{tab:barge_industrial}
\small
\setlength{\tabcolsep}{4.5pt}
\begin{tabular}{lcccc}
\toprule
Method & Hit@5 & Hit@10 & Hit@20 & Hit@50 \\
\midrule
\multicolumn{5}{l}{\textit{Scenario 1}} \\
GraphSAGE-based
& 0.2932
& 0.3743
& 0.4650
& 0.5951 \\
NANN
& 0.4416
& 0.4946
& 0.5636
& 0.6760 \\
OneRec
& \underline{0.5459}
& \underline{0.6132}
& \underline{0.6729}
& \underline{0.7348} \\
BARGE
& \textbf{0.6015}
& \textbf{0.6510}
& \textbf{0.6967}
& \textbf{0.7520} \\
\addlinespace[3pt]
\multicolumn{5}{l}{\textit{Scenario 2}} \\
GraphSAGE-based
& 0.0580
& 0.0895
& 0.1330
& 0.2064 \\
BERT-based
& 0.0523
& 0.0864
& 0.1347
& 0.2223 \\
OneRec
& \underline{0.0835}
& \underline{0.1256}
& \underline{0.1806}
& \underline{0.2658} \\
BARGE
& \textbf{0.0976}
& \textbf{0.1441}
& \textbf{0.2045}
& \textbf{0.2994} \\
\bottomrule
\end{tabular}
\end{table}

\subsubsection{Efficiency and Engineering Considerations}
\label{sec:barge_engineering}

BARGE's accuracy gains come at essentially zero marginal cost (Table~\ref{tab:barge_efficiency}). Using a 2-layer encoder instead of TIGER's 4-layer one absorbs the parameters of ICA, the HPR scorers, and the DPD dual towers, so BARGE is actually \emph{smaller} than TIGER ($19.91$\,M vs.\ $22.71$\,M), while per-epoch train/infer cost grows by under $10\%$ because the two DPD towers share the encoder and run in parallel.

\begin{table}[ht]
\centering
\caption{Efficiency comparison between TIGER and BARGE on Amazon Beauty under identical configurations.}
\label{tab:barge_efficiency}
\small
\begin{tabular}{l|ccc}
\toprule
Method & \#Params & Train (s/epoch) & Infer (s/epoch) \\
\midrule
TIGER & 22.71\,M & 22 & 17 \\
BARGE & \textbf{19.91\,M} & 24 & 18 \\
\bottomrule
\end{tabular}
\end{table}

Beyond headline cost, four engineering invariants and recipes were key to shipping BARGE:

\begin{enumerate}[leftmargin=1.4em,itemsep=2pt,topsep=2pt]
\item \textbf{Fixed evaluation budget.} Beam width stays at the TIGER-lineage default ($B = 20$) throughout offline evaluation, while the deployed retrieval channel runs $50$ SID sequences per request (\S\ref{sec:barge_deployment}); HPR reranks within the beam and OR-fusion truncates at the same $K$, so no downstream capacity re-planning was needed.
\item \textbf{Robust hyperparameter plateaus.} The HPR fusion weight follows an inverted-U with a mechanistic explanation (rescue-vs-damage trade-off), optimal at $\lambda = 0.25$; the reranker pool saturates beyond Top-$N \approx 400$. Both settings transferred across datasets without re-tuning.
\item \textbf{Cheaper codebook by design.} The layer-wise decreasing $(512, 256, 128, 64)$ codebook allocates capacity to the coarse layers where partitioning quality matters most, and uses \emph{fewer} codeword embeddings ($960$ vs.\ $1024$ for a uniform 4-layer configuration) while outperforming it.
\item \textbf{Orthogonality as a hard invariant.} The Householder parameterization enforces $R^\top R = I$ to $\sim 10^{-6}$ throughout training with no auxiliary loss, eliminating a class of regularization-tuning failure modes; the learned rotation reduces reconstruction loss by $0.04$--$0.05$ over freezing $R = I$.
\end{enumerate}

\subsubsection{Deployment and Online A/B}
\label{sec:barge_deployment}

\noindent\textbf{Online serving.} BARGE is integrated into Tencent's commercial media platform as a generative retrieval channel. For each request, BARGE uses beam search to generate $50$ SID sequences. Instead of mapping each SID sequence to a single item, we construct its query representation by summing the codebook embeddings corresponding to its constituent SID tokens, and use the resulting representation to retrieve relevant items from the item corpus through an embedding-based approximate nearest-neighbor retrieval service. Under the current per-request quota, each SID sequence retrieves $10$ items, producing at most $50\times 10=500$ candidates. These candidates are subsequently processed by the existing downstream cascade, including pre-ranking, fine-ranking, blending, filtering, and business-rule processing.

After cross-sequence deduplication, each request retains approximately $480$ distinct candidates on average, corresponding to a duplication rate of only $\approx\!4\%$. The limited overlap among the candidates retrieved
from the $50$ generated SID sequences indicates that the generation process does not collapse onto a small set of dominant items, but instead covers complementary regions of the item corpus. This candidate-level evidence suggests that BARGE maintains substantial retrieval diversity under a fixed beam budget.

\noindent\textbf{Online A/B test.} BARGE was trained on $2\times$ NVIDIA H20 GPUs and evaluated in a two-week online A/B test, in which $6\%$ of live traffic was allocated to the experimental group. All reported improvements are statistically significant. Compared with the deployed multi-stage incumbent, BARGE improved click-through rate by $+0.60\%$, the number of unique clicking users by $+1.34\%$, and total reading time by $+1.70\%$. 

Following the initial A/B test, we fully rolled out BARGE as a production retrieval channel and retained a long-term holdback group to continuously measure its incremental effect. Over the longer observation window, both click-through rate and total reading time exhibited larger and more stable improvements than those observed during the initial experimental period. One plausible explanation is the progressive reduction of the mismatch between the training and serving distributions. During the small-traffic experiment, many long-tail and heterogeneous items introduced by BARGE were under-represented in the historical exposure logs and could therefore be underestimated by downstream models. After the full rollout, the resulting exposure and user-feedback samples were gradually incorporated into the training pipeline, allowing the system to better model the item distribution introduced by BARGE and reducing the corresponding exposure bias.

The long-term holdback also captures cumulative system effects over a substantially longer period, while the larger observation window reduces the variance of the estimated treatment effect. Together, these factors help explain why the long-term gains are larger and more stable than the point estimates obtained from the short-term A/B test. The observation further indicates that the improvements introduced by BARGE are sustained after full production deployment.

\noindent\textbf{Channel-level analysis.} To further examine the contribution of BARGE at the retrieval stage, we deploy it as an individual retrieval channel and compare it with existing retrieval channels under the same traffic setting (Table~\ref{tab:recall_drilldown}). We report two metrics: the exposure adoption rate and CTR. The exposure adoption rate is defined as the ratio between the number of items from a retrieval channel that are eventually adopted and exposed by the downstream system and the total number of candidates submitted by that channel. It measures how effectively the retrieved candidates pass through downstream pre-ranking, fine-ranking, blending, filtering, and business-rule processing.

As shown in Table~\ref{tab:recall_drilldown}, BARGE achieves the highest CTR of $9.93\%$ among the four representative retrieval channels, compared with $8.70\%$ for graph-based item-to-item retrieval (Graph-based I2I), $6.94\%$ for DNN-based retrieval, and $5.61\%$ for content-based retrieval (CB). Its exposure adoption rate reaches $1.02\%$, which is comparable to the $1.03\%$ achieved by Graph-based I2I and places BARGE in the first tier among the evaluated retrieval channels. These results demonstrate BARGE's effectiveness along two complementary dimensions: its retrieved items maintain a competitive pass-through rate through the downstream cascade, while achieving the highest click conversion rate once exposed. Therefore, BARGE provides measurable standalone value when deployed as an individual retrieval channel.

\begin{table}[ht] 
\centering 
\caption{Online comparison of representative retrieval channels on Tencent's commercial media platform.} 
\label{tab:recall_drilldown} 
\begin{tabular}{lcc} 
\toprule 
Retrieval channel & Exposure adoption rate & CTR \\ 
\midrule 
Content-based (CB) & $0.47\%$ & $5.61\%$ \\ 
DNN-based & $0.85\%$ & $6.94\%$ \\ 
Graph-based I2I & $\mathbf{1.03\%}$ & $8.70\%$ \\ 
\textbf{BARGE} & $1.02\%$ & $\mathbf{9.93\%}$ \\ 
\bottomrule 
\end{tabular} 
\end{table}


\subsection{HiGR: Slate Generation for Industrial Recommendation}
\label{sec:higr_model}

\textbf{HiGR} (\textbf{Hi}erarchical \textbf{G}enerative slate \textbf{R}ecommendation)~\citep{higr_2026} is the
slate-generation model in TGR-GenRec. Instead of generating one item at a time, HiGR treats the \emph{entire ordered
slate as a single generation unit}. This formulation matches how recommendations are actually presented on our
surfaces: users consume an ordered list in one display, and its quality depends not only on the relevance of each
item but also on list-level ordering, compatibility, and diversity. Existing SID-based generative recommenders in
the OneRec lineage are designed for item-wise generation and therefore do not directly optimize these slate-level
properties. HiGR closes this gap through a co-designed pipeline comprising a \emph{prefix-contrastive tokenizer}, a
\emph{hierarchical slate decoder}, and \emph{listwise preference alignment}, which jointly structure the SID space,
generate slates efficiently, and align the resulting lists with holistic user preferences. Compared with OneRec under matched decoding settings (without KV caching), HiGR achieves more than a $5\times$
inference speedup, and retains the best quality--efficiency trade-off once KV caching is enabled
(\S\ref{sec:higr-efficiency}). HiGR has been deployed across multiple Tencent commercial
platforms, serving hundreds of millions of users and delivering consistent gains; the corresponding results are
reported in \S\ref{sec:higr-online}.

\subsubsection{Motivation: Three Disconnects Between Item Generation and Slate Optimization}
\label{sec:higr-motivation}
Discriminative two-stage systems score candidates in isolation and then assemble a slate by greedy selection,
heuristics, or reranking; optimizing items independently ignores inter-item dependencies and positional effects,
yielding slates that are locally reasonable but globally suboptimal. SID-based generative recommendation offers a more direct path: by generating an ordered slate autoregressively, it can jointly model item selection, ordering, and cross-item dependencies. However, extending item-wise SID generation to industrial-scale slate optimization exposes three disconnects:

\begin{itemize}
  \item \textbf{(D1) Entangled SID space.} Generative slate quality hinges on item tokenization, but standard
  residual quantization produces entangled, sparse code spaces in which SID \emph{prefixes} fail to reflect item
  semantics or collaborative relations. Inconsistent prefixes weaken controllability and make it hard to steer
  slate-level relevance and diversity through discrete codes.
  \item \textbf{(D2) Inefficient fine-grained decoding.} An $M$-item slate with $D$ SIDs per item requires decoding
  $M\times D$ tokens. Full-sequence autoregressive beam search over this length is slow under real-time constraints
  and entangles intra-item semantic composition with inter-item transitions, obscuring the global slate structure.
  \item \textbf{(D3) Objective misalignment.} Token-level likelihood does not optimize user-perceived slate quality.
  Platforms care about multiple list-level objectives---ranking fidelity, genuine user interest, and intra-list
  diversity---none of which is captured by next-token prediction alone.
\end{itemize}

\noindent HiGR addresses these three disconnects through a coordinated design spanning the tokenizer, decoder, and training
objective. PCRQ-VAE learns a structured hierarchical SID space for controllable slate generation (D1); HSD separates
global slate planning from local item decoding and confines beam search to short item-level sequences (D2); and
listwise ORPO aligns the generated slate with holistic user preferences (D3).

\begin{figure}[t]
\centering
\includegraphics[width=0.98\textwidth]{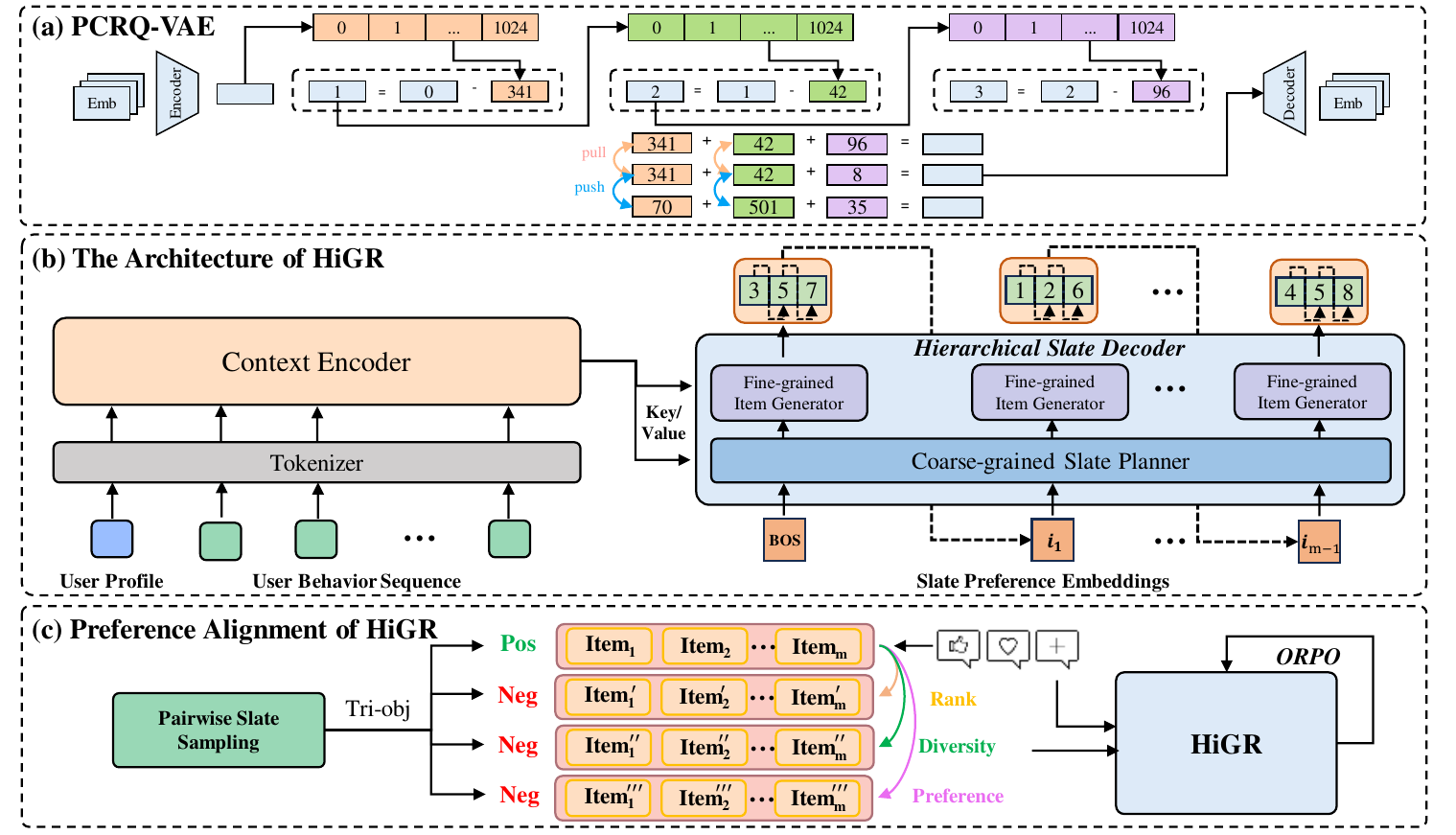}
\caption{Overall framework of HiGR: (a)~PCRQ-VAE for semantic tokenization with prefix-level contrastive alignment over the first $D{-}1$ codebook layers; (b)~Hierarchical Slate Decoder consisting of a coarse-grained slate planner (over preference embeddings) and a shared fine-grained item generator (over SIDs); (c)~ORPO-based listwise preference alignment over triple-objective slate preference pairs (rank fidelity, genuine interest, diversity).}
\label{fig:higr}
\end{figure}

\subsubsection{Method Overview}
\label{sec:higr-method}
As shown in Figure~\ref{fig:higr}, HiGR is a unified pipeline of three tightly coupled components---a
prefix-contrastive tokenizer, a hierarchical slate decoder, and listwise alignment---each acting on one of the three
disconnects. Given a user $u$ with history $S^{u}$, the model generates an $M$-item slate
$O^{u}=\mathcal{F}_{\theta}(u,S^{u})$ optimized for holistic list quality.

\paragraph{PCRQ-VAE --- structuring the discrete space (D1).}
The tokenizer keeps a standard RQ-VAE backbone but shapes the code space with two additions. A \emph{global
quantization} term aligns each item's aggregated codeword with its latent, preventing residual vanishing in deep
codebooks; and a \emph{prefix-contrastive} objective pulls the first $D{-}1$ codebook layers of semantically or
collaboratively related items together while excluding the leaf layer. High-level prefixes thus encode shared
semantics and the final layer preserves item identity, so slate decoding can impose relevance and diversity
constraints directly on discrete codes rather than on continuous embeddings.
\label{sec:higr_pcrqvae}

\paragraph{HSD --- coarse-to-fine generation (D2).}
Instead of autoregressing over the full $M\times D$ SID sequence, the hierarchical slate decoder factors generation
into a coarse \emph{slate planner} and a fine \emph{item generator}. The planner autoregresses in a compact
preference-embedding space---one embedding per slate position---to fix the global list intent, and a shared item
generator then grounds each preference into a short local SID sequence; an intra-list diversity regularizer on the
planned geometry keeps positions distinct. At inference the planner runs greedy while beam search is confined to the
short per-item decoding, shrinking the effective search space and cutting decoding cost substantially
versus full-sequence generation.
\label{sec:higr_hsd}

\paragraph{ORPO listwise alignment --- optimizing slate quality (D3).}
To connect token-level generation with user-perceived quality, HiGR aligns whole slates via reference-free Odds
Ratio Preference Optimization \citep{orpo_2024}, which folds preference optimization into supervised learning and needs no reference
model. Slate-level positive/negative pairs are mined from implicit feedback to cover three complementary objectives:
\emph{ranking fidelity} (against mis-ordered permutations), \emph{genuine interest} (against negatively received
items), and \emph{diversity} (against over-similar lists). The supervised term preserves generative accuracy while
the contrastive term rejects poorly ordered, irrelevant, or repetitive slates.

\subsubsection{Offline Results}
\label{sec:higr-offline}
\noindent\textbf{Public benchmark results.} We construct chronological user sequences on KuaiRec and adopt a
leave-five-out slate evaluation protocol. HiGR achieves the best result on every metric against traditional slate
recommendation (ListCVAE \citep{listcvae_2018}), discriminative sequential models (BERT4Rec and SASRec), and generative baselines (TIGER,
HSTU, and OneRec), as shown in Table~\ref{tab:higr-main}. Under the matched 25M-parameter setting, HiGR improves
NDCG@5 from $0.1496$ to $0.1574$ ($+5.2\%$) and Effective-View Recall@5 from $0.0422$ to $0.0491$ ($+16.4\%$) over
OneRec. Scaling HiGR to 100M parameters further raises these metrics to $0.1651$ and $0.0546$, corresponding to gains
of $10.4\%$ and $29.4\%$ over OneRec-25M, respectively.

\noindent\textbf{Industrial-scale offline evaluation.} The industrial dataset is collected from Tencent's commercial
media platform and contains one billion pre-training samples, of which $3\%$ are selected for ORPO post-training. We
evaluate both exposure modeling and positive-feedback quality: Impression Hit@5 and Recall@5 measure whether the
model recovers displayed items, Effective-View Hit@5 and Recall@5 focus on positively received items, and NDCG@5
measures their ordering quality. The compared systems span slate recommendation, discriminative sequential
recommendation, and SID-based generative recommendation, providing a broad test across modeling paradigms.

At the same 25M-parameter scale, HiGR outperforms the strongest baseline, OneRec-25M, on all five industrial metrics.
It improves Impression Hit@5 from $0.2438$ to $0.2825$ ($+15.9\%$), Effective-View Hit@5 from $0.1603$ to $0.1945$
($+21.3\%$), and NDCG@5 from $0.0589$ to $0.0714$ ($+21.2\%$). HiGR-25M without preference alignment already
surpasses OneRec-25M on all ten metrics across the two datasets, isolating gains from the tokenizer and hierarchical
decoder; ORPO then improves every metric further. The 100M model performs best overall, reaching $0.0831$ NDCG@5 on
the industrial dataset---a $41.1\%$ gain over OneRec-25M---and demonstrating that the improvements continue with
model capacity.

\begin{table}[t]
\centering
\caption{Offline performance on the industrial dataset and KuaiRec. Underlined and bold values denote the strongest
baseline and the best overall result, respectively.}
\label{tab:higr-main}
\small
\setlength{\tabcolsep}{4pt}
\resizebox{\linewidth}{!}{%
\begin{tabular}{l|ccccc|ccccc}
\toprule
\multirow{3}{*}{\textbf{Method}} & \multicolumn{5}{c|}{\textbf{Industrial dataset}} & \multicolumn{5}{c}{\textbf{KuaiRec}} \\
\cmidrule(lr){2-6} \cmidrule(lr){7-11}
& \multicolumn{2}{c}{Impressions} & \multicolumn{2}{c}{Effective Views} & \multirow{2}{*}{NDCG@5}
& \multicolumn{2}{c}{Impressions} & \multicolumn{2}{c}{Effective Views} & \multirow{2}{*}{NDCG@5} \\
\cmidrule(lr){2-3} \cmidrule(lr){4-5} \cmidrule(lr){7-8} \cmidrule(lr){9-10}
& Hit@5 & Recall@5 & Hit@5 & Recall@5 & & Hit@5 & Recall@5 & Hit@5 & Recall@5 & \\
\midrule
ListCVAE       & 0.0857 & 0.0178 & 0.0571 & 0.0117 & 0.0186 & 0.2304 & 0.0523 & 0.0757 & 0.0161 & 0.0630 \\
\midrule
BERT4Rec       & 0.0911 & 0.0187 & 0.0632 & 0.0129 & 0.0201 & 0.2612 & 0.0583 & 0.0829 & 0.0177 & 0.0713 \\
SASRec         & 0.1057 & 0.0218 & 0.0700 & 0.0143 & 0.0243 & 0.3017 & 0.0756 & 0.1028 & 0.0231 & 0.0839 \\
\midrule
TIGER          & 0.1812 & 0.0383 & 0.1204 & 0.0249 & 0.0406 & 0.4455 & 0.1299 & 0.1566 & 0.0378 & 0.1363 \\
HSTU           & 0.2281 & 0.0506 & 0.1487 & 0.0310 & 0.0492 & 0.5055 & 0.1334 & 0.1728 & 0.0397 & 0.1466 \\
OneRec-25M     & \underline{0.2438} & \underline{0.0577} & \underline{0.1603} & \underline{0.0367} & \underline{0.0589}
               & \underline{0.5150} & \underline{0.1395} & \underline{0.1782} & \underline{0.0422} & \underline{0.1496} \\
\midrule
HiGR-25M w/o ORPO & 0.2641 & 0.0612 & 0.1810 & 0.0395 & 0.0631 & 0.5218 & 0.1437 & 0.1805 & 0.0462 & 0.1536 \\
HiGR-25M        & 0.2825 & 0.0692 & 0.1945 & 0.0433 & 0.0714 & 0.5290 & 0.1485 & 0.1846 & 0.0491 & 0.1574 \\
\textbf{HiGR-100M}
                 & \textbf{0.3163} & \textbf{0.0760} & \textbf{0.2145} & \textbf{0.0495} & \textbf{0.0831}
                 & \textbf{0.5360} & \textbf{0.1544} & \textbf{0.1905} & \textbf{0.0546} & \textbf{0.1651} \\
\bottomrule
\end{tabular}%
}
\end{table}

\begin{figure}[t]
\centering
\includegraphics[width=0.85\textwidth]{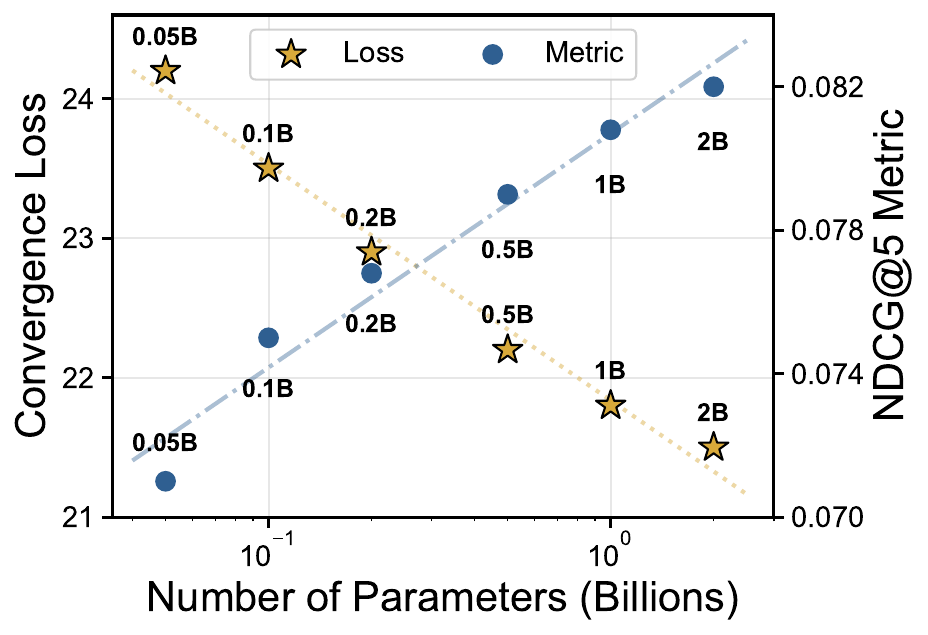}
\caption{Scaling behavior of HiGR from $0.05$B to $2$B parameters: training
loss and NDCG@5 follow power-law trends, indicating predictable returns to
model capacity for slate generation.}
\label{fig:higr_scaling}
\end{figure}

\noindent\textbf{Component analysis and scaling.} The ablations support the role of all three components. For
PCRQ-VAE, applying contrastive alignment to the first $D{-}1$ codebook layers yields $2.37\%$ collision, $93\%$
concentration, and $66.47\%$ consistency; extending the constraint to the leaf layer raises collision to $8.73\%$,
confirming the need to preserve the last layer for item identity. For HSD, removing the ILD regularizer reduces
intra-list diversity from $0.62$ to $0.59$ with nearly unchanged accuracy, whereas removing the context embedding
drops NDCG@5 from $0.0753$ to $0.0664$ (both measured in the ablation configuration of~\citep{higr_2026}, whose
pre-alignment base differs from the 25M main setting); separate item generators provide no consistent gain over the
shared design.
For preference alignment, ORPO achieves ILD scores of $0.62$ and $0.63$ on the industrial dataset and KuaiRec,
respectively, compared with $0.57$ and $0.60$ for DPO \citep{dpo_2023}. Ranking fidelity and genuine-interest objectives primarily
improve NDCG@5, while the diversity objective primarily improves ILD; jointly optimizing all three produces the best
NDCG@5 of $0.0831$ while retaining an ILD of $0.62$. Finally, experiments from 0.05B to 2B parameters follow a clear
power-law trend in loss and NDCG@5 (Figure~\ref{fig:higr_scaling}), indicating predictable gains from scaling model
capacity---extending the LLM-style scaling behavior observed for generative ranking (\S\ref{sec:genrank_scaling})
to whole-slate generation. The complete ablation
tables and hyperparameter analyses are available in the original HiGR paper~\citep{higr_2026}.

\subsubsection{Efficiency and Engineering Considerations}
\label{sec:higr-efficiency}

\noindent\textbf{Efficiency from hierarchical slate decoding.} A flat slate generator represents an $M$-item slate
as one sequence of $M D$ SID tokens, where each item contains $D$ codewords. This couples global slate modeling with
fine-grained item construction and forces both training and inference to operate on a long target sequence. HSD
factorizes the computation into an $M$-step slate planner over preference embeddings and a shared item generator over
short $D$-token SID sequences. During training, this replaces target-side self-attention over one $M D$-token sequence
with attention over the $M$ planner positions plus a batch of short item-level sequences. The same item generator is
shared across all slate positions, so its parameter count does not grow with slate length.
HSD therefore improves the training-side computational structure and parameter efficiency, although the paper does
not report a standalone training-speed measurement.

\begin{figure}[t]
\centering
\includegraphics[width=0.6\textwidth]{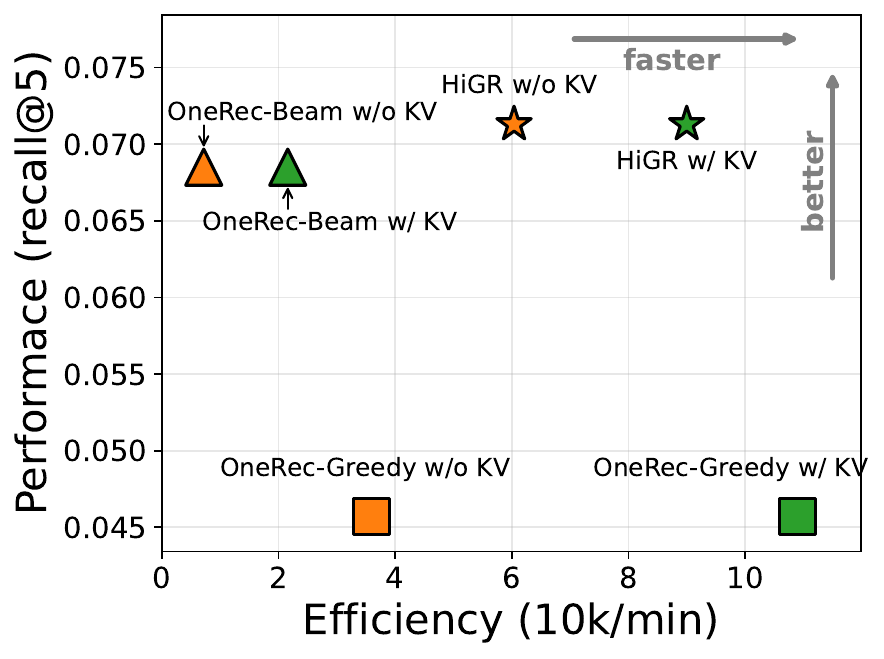}
\caption{Quality--efficiency trade-off of HiGR against OneRec-style decoding (beam search and greedy), with and without KV caching. HiGR attains the best Recall@5 at substantially higher decoding throughput.}
\label{fig:higr_efficiency}
\end{figure}

At inference, the efficiency gain is more direct. A OneRec-style decoder carries beam search through the entire
$M D$-token slate, whereas HSD performs greedy autoregressive planning over only $M$ preference embeddings and
confines beam search to each short $D$-token item sequence. Its decoding complexity is reduced from
$\mathcal{O}(B M^{3}D^{3}l_{\mathrm{slate}}d)$ to
$\mathcal{O}(M^{3}l_{\mathrm{slate}}d+B M D^{3}l_{\mathrm{item}}d)$, where $B$ is the beam width, $d$ the hidden dimension, and $l_{\mathrm{slate}}$ and $l_{\mathrm{item}}$ the depths of the planner and item generator (see Appendix~\ref{app:complexity}). Under a fixed total depth, allocating $14$ layers to the planner and $2$ to the item generator provides the best production
trade-off: compared with a $12{:}4$ split, it lowers latency by $26\%$ with only a $0.6\%$ relative decrease in
NDCG@5. Without KV caching, HiGR achieves more than a $5\times$ inference speedup over OneRec-Beam (OneRec-style
full-sequence decoding with beam search) while improving Recall@5 by over $5\%$; with KV caching, it retains the best quality--efficiency trade-off (Figure~\ref{fig:higr_efficiency})~\citep{higr_2026}. In production serving, this decoding structure holds end-to-end P99 latency below $50$\,ms while reducing GPU demand by approximately $60\%$ relative to a OneRec-style full-sequence autoregressive baseline at equivalent throughput---the serving envelope quoted in Table~\ref{tab:genrec_models}.

\noindent\textbf{Beam-search optimization.} We further optimize the local SID beam search through a
Flash-Attention-aware tensor layout. At the first SID level, all beams share the same preference embedding, so the
item generator runs only once and its top-$B$ codewords initialize the beams, avoiding $B$ identical decoder calls.
At subsequent levels, all beam prefixes are decoded in one batched forward pass; cumulative scores over the
$B\times V$ beam--codeword candidates are then flattened and pruned with a single top-$B$ operation per request.

The main memory optimization comes from using two sequence layouts for the same flattened decoder states. For causal
self-attention, they are interpreted as $N B$ independent beam sequences, preserving isolation between beams, where
$N$ is the request batch size. For cross-attention, the $B$ beams of each request are packed into one query sequence
and attend to a single ragged user-history representation. This removes padding from variable-length histories and,
more importantly, avoids replicating the encoder memory once per beam. The implementation therefore preserves exact
beam-search scoring while reducing redundant first-step computation, Python-side beam iteration, and beam-dependent
encoder-memory materialization.

\subsubsection{Deployment and Online A/B}
\label{sec:higr-online}

HiGR has been deployed and evaluated through online A/B experiments in three Tencent commercial recommendation
scenarios. Table~\ref{tab:higr-online} reports the relative lift over the corresponding incumbent production system
in each scenario. Scenario~1 ($5\%$ live traffic) shows consistent engagement gains: Average Stay Time, Average Watch Time, Average Video
Views, and Average Request Count increase by $1.03\%$, $1.22\%$, $1.73\%$, and $1.57\%$, respectively. Scenario~2
improves both user response and monetization, raising CTR by $0.68\%$ and advertising revenue by $0.56\%$.
Scenario~3 further improves Average Watch Time by $1.14\%$ and Average Video Views by $0.88\%$. The gains across all
three scenarios show that HiGR transfers beyond a single deployment setting and consistently improves engagement,
traffic consumption, and commercial value.

\begin{table}[t]
\centering
\caption{Relative lift from online A/B experiments across three Tencent commercial recommendation scenarios.}
\label{tab:higr-online}
\small
\setlength{\tabcolsep}{5pt}
\begin{tabular}{l|cccc}
\toprule
\multicolumn{5}{c}{\textbf{Scenario 1}} \\
\midrule
Metric
& \shortstack{Average\\Stay Time}
& \shortstack{Average\\Watch Time}
& \shortstack{Average\\Video Views}
& \shortstack{Average\\Request Count} \\
Lift (\%) & 1.03 & 1.22 & 1.73 & 1.57 \\
\midrule
\multicolumn{5}{c}{\textbf{Scenario 2}} \\
\midrule
Metric & \multicolumn{2}{c}{CTR} & \multicolumn{2}{c}{Ad Revenue} \\
Lift (\%) & \multicolumn{2}{c}{0.68} & \multicolumn{2}{c}{0.56} \\
\midrule
\multicolumn{5}{c}{\textbf{Scenario 3}} \\
\midrule
Metric & \multicolumn{2}{c}{Average Watch Time} & \multicolumn{2}{c}{Average Video Views} \\
Lift (\%) & \multicolumn{2}{c}{1.14} & \multicolumn{2}{c}{0.88} \\
\bottomrule
\end{tabular}
\end{table}

\newif\ifshowowner
\showownerfalse
\newcommand{\owner}[1]{\ifshowowner{\textcolor{blue}{\textbf{(#1)}}\ }\fi}
\section{TGR-Reason: Reasoning-Augmented Generative Recommendation}
\label{sec:think}

\textbf{TGR-Reason} is the reasoning-augmented direction of the TGR Stack. It turns LLM-derived intent inference into reasoning-enriched semantic IDs (\emph{reason tokens}) that connect offline reasoning, memory retrieval, and online generative decision-making. A Think model trained with LatentRec \citep{latentrec} generates, outside the request path, a user's top-$K_r$ candidate items as complete SIDs. At serving time, these tokens follow two complementary paths. They directly provide semantic prompts to TGR-GenRec (\S\ref{sec:higr}) and act as queries for Reasoning-Guided Group Memory Retrieval (GMR), which retrieves relevant behavioral evidence from Group Memory. The retrieved group context is combined with the user's Personal Memory---the observed behavior sequence---before being consumed by the generator. This design introduces no online reasoning rollout or additional autoregressive decision, and targets the requests for which co-occurrence provides the least evidence: cold-start, long-tail, and ambiguous-intent cases. In the deployed direct-injection configuration on one of Tencent's commercial content platforms, reason tokens raise cold-start new-user Hit@1 from $0.0451$ to $0.2606$ ($+477.8\%$), with consistent improvements across every user cohort (\S\ref{sec:reason_results}). All three modules operate in the shared SID space and reuse the tokenizer, encoder, and decoder interfaces of TGR-GenRec (principle~P2, \S\ref{sec:tgr_stack}).

\subsection{Motivation}
\label{sec:reason_motivation}

The TGR-GenRec generator is trained to reproduce feedback-weighted co-occurrence. This works well when a user's history densely constrains the next item. It falls short on the requests that matter most: new users with almost no history, long-tail items with sparse collaborative signal, and sessions of ambiguous intent. Such requests are underdetermined by co-occurrence alone. They call for external world knowledge and for the intent inference that a pattern-matching model never performs, and they are precisely where reasoning has the most to contribute.

This limitation also exposes a memory imbalance. A user's observed sequence forms a precise but incomplete \emph{Personal Memory}: it reflects individual interest, yet becomes uninformative when the history is short or ambiguous. Population-wide interaction sequences form a much richer \emph{Group Memory}, but retrieving from them without an intent signal introduces popular yet irrelevant behavior. Reason tokens bridge the two. They infer plausible interests beyond the available Personal Memory and serve as semantic queries that select intent-consistent evidence from Group Memory, allowing population behavior to supplement rather than overwrite the individual's own history.

Two lines of work make an LLM recommender reason, and both run it on the request path. Explicit reasoning writes the reasoning as language. OneRec-Think \citep{onerec_think_2025}, for instance, produces a textual chain of thought before decoding each SID. This is interpretable, but it emits hundreds of extra tokens per request. Latent reasoning instead reasons in the hidden-state space. ReaRec \citep{rearec_2025} rolls the sequence representation forward for several steps before prediction, which removes the token cost but still runs the rollout at inference; STREAM-Rec \citep{stream_rec_2025} similarly moves sequential recommendation toward multi-step slow thinking at inference. Both regimes tie reasoning capacity to per-request computation, which industrial latency budgets cannot absorb. TGR-Reason moves reasoning off the request path entirely. Its Think model builds on the LatentRec framework \citep{latentrec}, installs reasoning through training-time supervision, and periodically produces reason tokens in an offline pipeline. At serving time no LLM reasoning rollout is executed, so the reasoning cost is amortized rather than paid on every impression. To our knowledge, TGR-Reason is among the first to bring training-time-amortized reasoning to an industrial generative recommender \citep{fm4recsys_2025}.

\subsection{Method Overview}
\label{sec:reason_decoding}

\begin{figure}[t]
\centering
\includegraphics[width=1\textwidth]{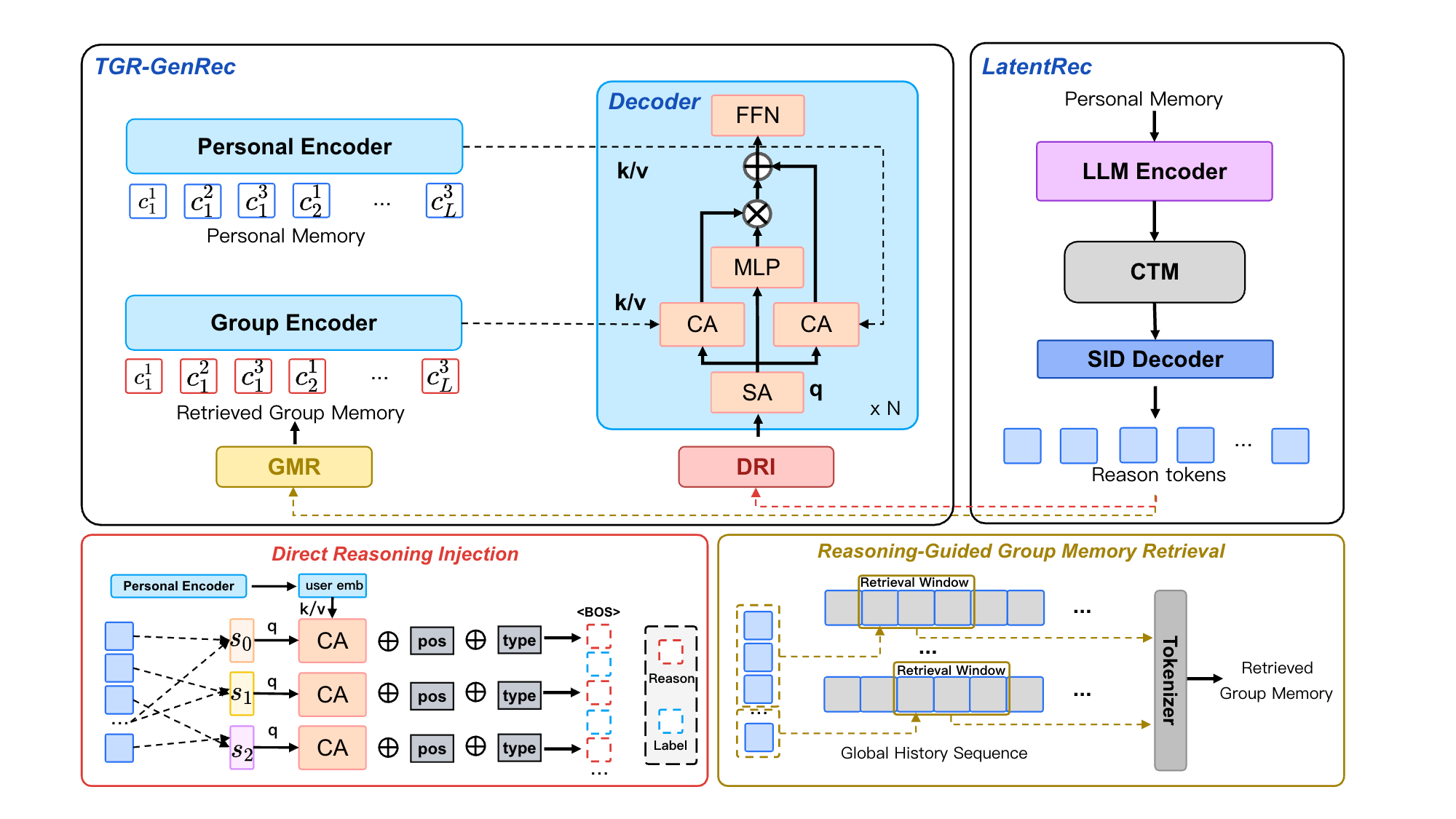}
\caption{TGR-Reason and memory augmentation in the middle layer of the TGR Stack. An offline reasoner produces reasoning-enriched SIDs that directly guide TGR-GenRec decoding through DRI and serve as token-level queries for GMR. TGR-Memory combines the GMR-retrieved collaborative evidence with the user's Personal Memory and injects the resulting dual-memory context into TGR-GenRec.}
\label{fig:reason_framework}
\end{figure}

Figure~\ref{fig:reason_framework} organizes the framework into three cooperating modules. \emph{TGR-Reason} generates reason tokens in an offline pipeline. Its Think model is built on LatentRec \citep{latentrec}: per-step supervision installs reasoning into the model parameters, allowing the deployed model to emit top-$K_r$ complete SIDs through a standard decoding pass without an inference-time reasoning rollout. \emph{TGR-Memory} maintains two complementary sources. Personal Memory is derived from the current user's observed behavior and preserves individual interest; Group Memory indexes population-wide interaction sequences, from which GMR selects intent-consistent collaborative evidence using reason-token queries. \emph{TGR-GenRec} consumes these signals through two interfaces: Direct Reasoning Injection (DRI) transforms reason tokens into level-aligned decoder representations, while GMR supplies retrieved Group Memory that is encoded and adaptively fused with Personal Memory. The two paths preserve the original number of generated SID tokens and introduce no additional autoregressive reasoning step. Group Memory retrieval and encoding remain bounded, non-autoregressive operations.

\subsubsection{Reason Tokens as a Shared Reasoning--Memory Interface}
\label{sec:reason_taxonomy}

A generative recommender can externalize reasoning in several forms, and that choice determines both how the signal enters the system and what must run on the request path. Table~\ref{tab:reason_taxonomy} contrasts four alternatives. Chain-of-thought text exposes the reasoning process but requires long autoregressive generation. Auxiliary generated features \citep{recgpt_2025} introduce a separate representation and serving interface. Latent reasoning states avoid text generation, but remain internal to the model and cannot directly connect independently maintained reasoning, memory, and recommendation modules. Reason tokens instead express the Think model's output as complete SIDs in the same vocabulary used throughout the TGR Stack.

This shared representation gives each reason token two complementary roles. As a \emph{reasoning carrier}, its layer-wise SID evidence directly prompts TGR-GenRec decoding. As a \emph{memory query}, the complete SID drives GMR to retrieve intent-consistent behavior from Group Memory, which is subsequently combined with Personal Memory. No natural-language rationale crosses the module boundary, and no LLM reasoning rollout is added to online serving. The memory path incurs bounded retrieval and encoding work, but both paths preserve the existing SID tokenizer and autoregressive output space of \S\ref{sec:tgr_stack}. Latent reasoning remains an internal training mechanism, and the reason tokens are the primary interface shared across TGR-Reason, TGR-Memory, and TGR-GenRec.

\begin{table}[t]
\centering
\caption{Alternative carriers for connecting reasoning with generative recommendation. Latent states are used internally during Think-model training, whereas offline-generated reason tokens form the shared interface to memory and online decoding.}
\label{tab:reason_taxonomy}
\small
\setlength{\tabcolsep}{5pt}
\begin{tabular}{l|l|l|l}
\toprule
Carrier & Produced & Request-path cost & Role in TGR-Reason \\
\midrule
Chain-of-thought text       & Inference time   & High AR cost          & Not adopted \\
Auxiliary generated features & Offline / online & Extra feature stack   & Not used \\
Latent reasoning state      & Training time    & None at $K{=}0$       & Internal supervision \\
Reason tokens (SIDs)        & Offline          & No reasoning rollout  & Shared interface \\
\bottomrule
\end{tabular}
\end{table}

\subsubsection{Offline Reason Token Generation with LatentRec}
\label{sec:reason_latent}

\begin{figure}[t]
\centering
\includegraphics[width=0.8\linewidth]{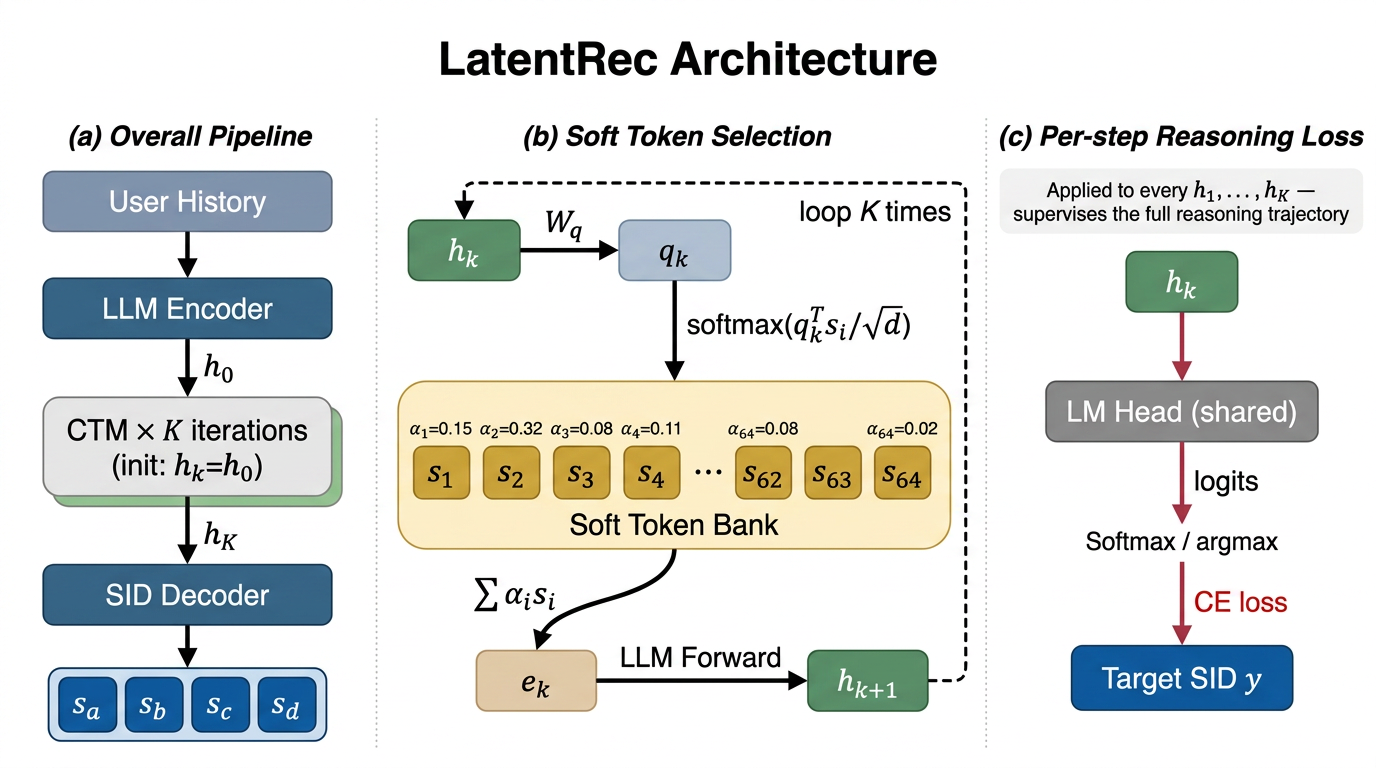}
\caption{LatentRec architecture inside the Think model. From the user-context hidden state $\mathbf{h}_0$, the Continuous Thought Module unrolls $K$ latent steps. Soft Token Selection projects each hidden state back into the input-embedding distribution before it re-enters the backbone, and the Per-step Reasoning Loss (Eq.~\ref{eq:reason_prl}) supervises every intermediate state through the shared prediction head against the first level of the target semantic ID. The reasoning module runs only during training and is skipped at $K{=}0$ offline generation.}
\label{fig:reason_latentrec}
\end{figure}

The Think model is the offline reason-token generator of TGR-Reason, implemented as an SID-based generative recommender trained with our LatentRec framework \citep{latentrec}. LatentRec replaces inference-time reasoning, whether verbalized as chain-of-thought text or rolled out as latent states, with supervision applied only during training. The resulting model retains the reasoning gain while generating reason tokens through an ordinary SID-decoding pass in the periodic offline pipeline. Three components make this amortization possible: the Continuous Thought Module, Soft Token Selection (STS), and Per-step Reasoning Loss (PRL) (Figure~\ref{fig:reason_latentrec}).

\paragraph{Continuous Thought Module.} The module produces the intermediate hidden states that per-step supervision later scores. Starting from the user-history hidden state $\mathbf{h}_0$, taken as the last-layer representation of the final history token, it unrolls $K$ latent reasoning steps,
\begin{equation}
\mathbf{e}_k = \phi(\mathbf{h}_k), \qquad \mathbf{h}_{k+1} = \mathrm{LLM}(\mathbf{e}_k, \mathrm{KV}^{(k)}),
\end{equation}
where $\phi$ is the Soft Token Selection projection defined below and $\mathrm{KV}^{(k)}$ is the key-value cache accumulated over the history tokens and the first $k$ latent steps. Each latent step is a single-token forward pass that attends to this growing cache, so it reuses the history computation rather than recomputing it, and the rollout adds $K$ such passes during training. The recurrence is deliberately minimal. It introduces no reasoning-specific Transformer weights, and its only trainable additions are the Soft Token Selection parameters, about $2.5$\,M in total and under $0.2\%$ of the backbone. Because the rollout feeds back through the same backbone under the same key-value cache, disabling it is exact rather than approximate. Setting $K{=}0$ during offline reason-token generation skips the module entirely and decodes directly from $\mathbf{h}_0$, which is the production configuration used here.

\paragraph{Soft Token Selection.} A representation-space mismatch makes naive latent rollout unstable in the SID-decoding regime. In the Qwen3-1.7B backbone the last-layer hidden states carry roughly $14\times$ the L2 norm of the input embeddings. Feeding a raw hidden state back as the next input therefore drives the per-step predictions off the SID vocabulary as $K$ grows, and the problem is sharper here than in math or logic latent reasoning because every intermediate state is also scored through the prediction head and must stay decodable. STS resolves this by never feeding a hidden state back directly. It maintains $N{=}64$ learnable soft tokens $\mathbf{S}$ initialized from the vocabulary embeddings and re-expresses each hidden state as a convex combination of them,
\begin{equation}
\alpha_i^{(k)} = \frac{\exp(\mathbf{q}_k^\top \mathbf{s}_i / \sqrt{d})}{\sum_j \exp(\mathbf{q}_k^\top \mathbf{s}_j / \sqrt{d})}, \qquad \mathbf{e}_k = \sum_i \alpha_i^{(k)} \mathbf{s}_i,
\end{equation}
with query $\mathbf{q}_k = W_q \mathbf{h}_k$. Since $\mathbf{e}_k$ is a convex combination of vocabulary-scale vectors, it stays within the input-embedding manifold in both norm and direction, whereas a scalar rescaling would fix the norm but leave the direction off-manifold. This keeps every projected input decodable by the shared prediction head, which is what lets the same head supervise all $K$ steps. Empirically, the distinction is not cosmetic. Under matched settings, raw hidden-state feedback combined with per-step head supervision diverges within the first training epoch as the intermediate logits leave the SID vocabulary, so STS is necessary in this regime rather than a convenience.

\paragraph{Per-step Reasoning Loss.} The third component supervises every intermediate state $\mathbf{h}_k$ through the shared prediction head, and it is the component that carries most of the gain. It adapts the progressive supervision of ReaRec \citep{rearec_2025}, introduced for sequential-recommendation encoders, to the LLM SID-decoder regime by applying the shared prediction head at every latent step under a temperature that sharpens toward the last step,
\begin{equation}
\mathcal{L}_{\text{PRL}} = \frac{1}{K}\sum_{k=1}^{K} \mathrm{CE}\!\left(\mathrm{LM\_head}(\mathbf{h}_k)\,/\,T_k,\; y\right), \qquad T_k = T_b + T_s\!\left(1 - \tfrac{k}{K}\right),
\label{eq:reason_prl}
\end{equation}
where $y$ is the first level of the target semantic ID and $T_k$ decays linearly from a relaxed temperature at the first step to $T_b$ at step $K$. The full objective is $\mathcal{L} = \mathcal{L}_{\text{CE}} + \lambda\,\mathcal{L}_{\text{PRL}}$ with $\lambda{=}0.1$. Supervising the first SID level gives the intermediate steps a stable target, since that level carries the coarsest routing decision in the RQ codebook. Supervising deeper levels would instead require teacher-forcing the target SID prefix at each step, which couples the latent rollout to a fixed decoding path and conflicts with the autoregressive SID decoding used at inference, so those levels are left to $\mathcal{L}_{\text{CE}}$. Read this way, PRL is multi-step deep supervision through the prediction head, which is consistent with its gain surviving at $K{=}0$.

The property that makes reasoning amortizable is that this supervision improves accuracy even when generation uses $K{=}0$, where the module never runs in the forward pass. The gain lives in the LoRA adapters shaped by per-step supervision rather than in multi-step computation at generation time, so the reasoning module is discarded after training and the offline pipeline reduces to standard SID decoding. A chain-of-thought-distilled initialization is optional. It is not required at $K{=}0$, but it stabilizes deeper rollout, placing the value of multi-step latent computation at training time rather than on the production generation path. The production Think model therefore generates reason tokens with $K{=}0$.

\begin{figure}[t]
\centering
\includegraphics[width=\textwidth]{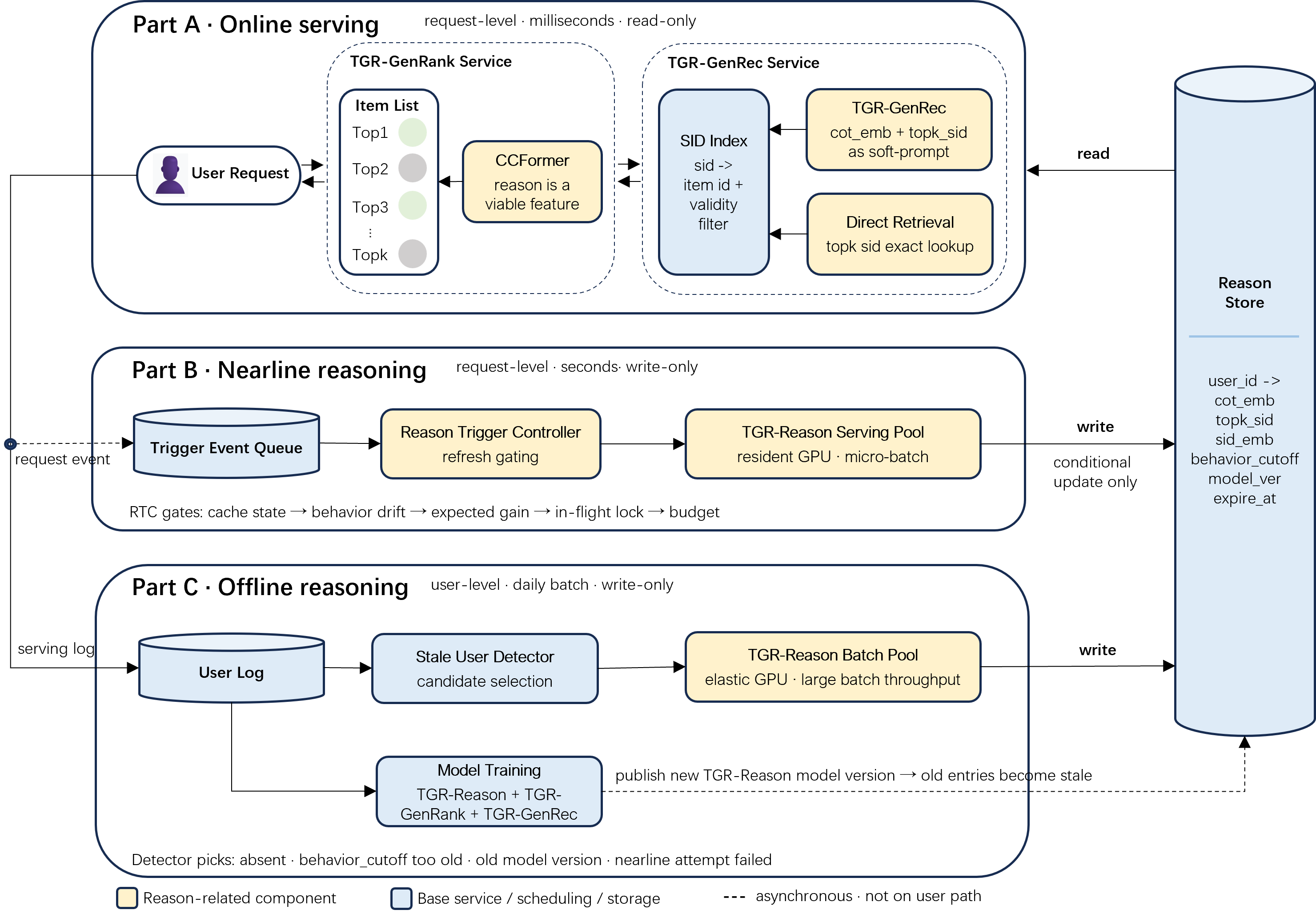}
\caption{The three-tier deployment architecture of TGR-Reason. 
\textbf{Part A: Online serving} operates at request level under a millisecond latency budget. It reads a pre-materialized Reason record from the Reason Store and uses the record for both reason-enhanced generation and direct SID retrieval.
\textbf{Part B: Nearline reasoning} is asynchronously triggered by online requests. The Reason Trigger Controller (RTC) admits only refreshes with sufficient expected utility, which are executed by a resident-GPU serving pool and conditionally written back to the store.
\textbf{Part C: Offline reasoning} periodically detects missing, stale, failed, or version-incompatible users and regenerates their Reason records using a throughput-oriented batch pool; a separate scheduled job retrains and publishes the associated models.
The online tier is read-only, whereas the nearline and offline tiers update the shared Reason Store through version-aware conditional writes.
}
\label{fig:reason_deployment}
\end{figure}

\paragraph{Reason tokens.} For each user refresh, the Think model outputs its top-$K_r$ predicted items, where $K_r$ denotes the number of retained reasoning candidates and each prediction is a complete $L$-level semantic ID. These \emph{reason tokens} are the primary artifact that TGR-Reason exports: they directly guide TGR-GenRec decoding through DRI and serve as semantic queries for GMR. The Think model can additionally export a compact embedding that summarizes the reasoning behind each prediction, shown as the \texttt{cot\_emb} feature in Figure~\ref{fig:reason_deployment}, which the generator may consume as an auxiliary soft prompt alongside the SIDs. This embedding is precomputed during the offline refresh, so the request path still produces no natural-language reasoning, and the deployed configuration behind our results relies on the semantic-ID reason tokens alone. The notation $K_r$ distinguishes the number of exported reason tokens from the $K$ latent reasoning steps used only during LatentRec training.

\subsubsection{Reasoning and Memory Injection into TGR-GenRec}
\label{sec:reason_inject}

The downstream generator reuses the shared SID tokenizer and encoder--decoder backbone of TGR-GenRec. For user $u$, let $S^u$ denote the historical SID sequence that forms Personal Memory, and let the Think model provide $K_r$ reason tokens, $\mathcal{R}_u=\{\mathbf{r}_{u,1},\ldots,\mathbf{r}_{u,K_r}\}$, where $\mathbf{r}_{u,k}=(r_{u,k}^{1},\ldots,r_{u,k}^{L})$ is a complete $L$-level SID. As shown in Figure~\ref{fig:reason_framework}, these tokens enter TGR-GenRec through two complementary interfaces. DRI transforms their layer-wise semantic evidence into Personal-Memory-conditioned representations aligned with SID decoding. GMR instead uses the complete reason tokens to query Group Memory; the retrieved population evidence is then encoded and adaptively fused with Personal Memory.

\paragraph{Direct Reasoning Injection.} DRI converts the $K_r$ reason tokens into Personal-Memory-conditioned, layer-wise representations for hierarchical SID decoding. A single reason token describes a coherent candidate item, but its codewords play different roles along the coarse-to-fine SID hierarchy. DRI therefore reorganizes the reason-token set by codebook level while retaining all $K_r$ alternatives. For level $l$, it looks up the shared SID embeddings
\begin{equation}
\mathbf{E}_{u}^{l}
=
[\mathbf{e}_{l}(r_{u,1}^{l}),\ldots,\mathbf{e}_{l}(r_{u,K_r}^{l})]
\in\mathbb{R}^{K_r\times d},
\end{equation}
where $\mathbf{e}_{l}$ is the embedding table of the $l$-th codebook and $d$ is the SID embedding dimension. No rank or candidate-position embedding is added: the reason tokens constitute an unordered evidence set rather than a reasoning sequence.

The relevance of this reasoning evidence depends on Personal Memory. Let $\mathbf{X}_u$ denote the $N_u$ encoder-side context embeddings $[\mathbf{x}_{u,1},\ldots,\mathbf{x}_{u,N_u}]$ derived from the user's profile and historical SID sequence, and let $m_{u,i}$ indicate a non-padding position. The block summarizes Personal Memory with masked mean pooling,
\begin{equation}
\mathbf{q}_{u}
=
\frac{\sum_{i=1}^{N_u}m_{u,i}\mathbf{x}_{u,i}}
     {\max(1,\sum_{i=1}^{N_u}m_{u,i})}.
\end{equation}
The same Personal-Memory query is then matched against the $K_r$ reason-token codewords independently at each injected level. In particular, the $l$-th level has its own multi-head cross-attention parameters, with $H$ heads of dimension $d_h$:
\begin{align}
\mathbf{o}_{u,h}^{l}
&=
\operatorname{softmax}\!\left(
\frac{(\mathbf{q}_{u}\mathbf{W}_{Q,h}^{l})
      (\mathbf{E}_{u}^{l}\mathbf{W}_{K,h}^{l})^{\top}}
     {\sqrt{d_h}}+\mathbf{A}_{u}
\right)
(\mathbf{E}_{u}^{l}\mathbf{W}_{V,h}^{l}),\\
\bar{\mathbf{z}}_{u}^{l}
&=
\operatorname{Concat}_{h=1}^{H}(\mathbf{o}_{u,h}^{l})\mathbf{W}_{O}^{l},
\qquad
\mathbf{z}_{u}^{l}=\bar{\mathbf{z}}_{u}^{l}+\mathbf{p}_{0}+\boldsymbol{\tau}_{l}.
\end{align}
Here, $\mathbf{A}_{u}$ is the additive attention mask over the $K_r$ reason tokens (zero for available SIDs and $-\infty$ for missing ones), $\mathbf{p}_{0}$ is the item-position embedding, and $\boldsymbol{\tau}_{l}$ is the codebook-level type embedding. Because no candidate-order encoding is used, permuting the reason tokens leaves $\mathbf{z}_{u}^{l}$ unchanged. More importantly, the attention weights are conditioned on $\mathbf{q}_{u}$: rather than copying the top-ranked LLM prediction, the block learns which reasoning candidates are compatible with Personal Memory. Separate attention parameters across levels further allow coarse intent codes and fine-grained residual codes to select different evidence from the same reason-token set. If all reason tokens are missing, the resulting prompt positions are masked from the decoder.

The fused vectors $\mathbf{Z}_u=[\mathbf{z}_u^{1},\ldots,\mathbf{z}_u^{L_t}]$ act as continuous, hierarchy-aligned prompts, where $L_t\leq L$ is the number of injected levels. Let $y^l$ denote the target item's codeword at SID level $l$. The prompts are interleaved immediately before their corresponding target codewords:
\begin{equation}
[\mathrm{BOS}\mid \mathbf{z}_u^{1},y^{1}\mid\cdots\mid
\mathbf{z}_u^{L_t},y^{L_t}\mid y^{L_t+1},\ldots,y^{L}].
\end{equation}
A causal mask ensures that the decoder state at $\mathbf{z}_u^{l}$ predicts $y^{l}$ without observing that target codeword; for levels beyond $L_t$, prediction follows standard autoregressive SID decoding. Meanwhile, every decoder state still cross-attends to Personal Memory, so the prompts augment rather than replace the user's observed behavior. During beam search, $\mathbf{Z}_u$ is computed once before decoding and shared by all expanded beams. Consequently, direct reasoning injection exploits all $K_r$ reason tokens without generating them online, changing neither the number of SID decisions nor the beam-search space.

\paragraph{Reasoning-Guided Group Memory Retrieval (GMR).} Group Memory is an indexed corpus of population-wide behavior sequences, and each complete reason token acts as a semantic retrieval query over that memory. This complements DRI: DRI transfers the Think model's inferred intent into decoding, while GMR grounds that intent in what users collectively consume afterward. Formally, let $\mathcal{G}=\{\mathbf{G}^{(v)}\}_{v}$ denote Group Memory, where $\mathbf{G}^{(v)}=[\mathbf{g}_{1}^{(v)},\ldots,\mathbf{g}_{T_v}^{(v)}]$ is user $v$'s interaction sequence and every $\mathbf{g}_{t}^{(v)}$ is a complete $L$-level SID. For reason token $\mathbf{r}_{u,k}$, token-guided retrieval locates its occurrences across Group Memory and pools the $W$ items immediately following each occurrence:
\begin{equation}
\mathcal{B}_{u,k}
=\biguplus_{(v,t):\,\mathbf{g}_{t}^{(v)}=\mathbf{r}_{u,k}}
\left\{\mathbf{g}_{t+j}^{(v)}\;\middle|\;
1\leq j\leq W,\;t+j\leq T_v\right\},
\end{equation}
where $\biguplus$ denotes multiset union, retaining repeated successor occurrences across users and sequence positions for frequency counting. For each distinct successor SID, GMR counts its occurrences in the pooled multiset, ranks all successors by this frequency, and independently retains the $M$ most frequent ones for query $\mathbf{r}_{u,k}$. We denote the resulting ranked list by $\mathcal{C}_{u,k}=[\mathbf{c}_{k,1},\ldots,\mathbf{c}_{k,M}]$.
Applying the same operation to all $K_r$ reason tokens produces $K_r$ query-specific successor lists, which are concatenated rather than globally reranked:
\begin{equation}
\mathcal{C}_u
=\mathcal{C}_{u,1}\Vert\cdots\Vert\mathcal{C}_{u,K_r},
\qquad |\mathcal{C}_u|=K_r M.
\end{equation}
Processing each reason token independently preserves the downstream behavior associated with every inferred intent: the forward window captures what users tend to consume after that SID, and the per-query top-$M$ selection ensures that each reason token contributes $M$ frequently observed subsequent items to the retrieved Group Memory context.

\paragraph{Dual-Memory Encoding and Adaptive Fusion.} Each retrieved item $\mathbf{c}_{k,m}$ is represented by the same $L$-level SID tokenizer used for Personal Memory. Separate encoders then construct a Group Memory representation from the concatenated $K_r M$ retrieved items and a Personal Memory representation from the user's observed history,
\begin{equation}
\mathbf{H}_u^{g}
=\operatorname{Enc}_{\mathrm{group}}
\!\left(\operatorname{Tok}_{\mathrm{SID}}(\mathcal{C}_u)\right),
\qquad
\mathbf{H}_u^{p}
=\operatorname{Enc}_{\mathrm{personal}}
\!\left(\operatorname{Tok}_{\mathrm{SID}}(S^u)\right).
\end{equation}
The shared SID space makes the two memories semantically compatible, while separate encoders preserve their different provenance: $\mathbf{H}_u^{p}$ records the current user's individual interest, whereas $\mathbf{H}_u^{g}$ summarizes population-level successor behavior selected by the reason tokens.

At SID decoding level $l$, the self-attention state $\mathbf{d}_l$ queries Personal Memory and Group Memory through separate cross-attention operators,
\begin{equation}
\mathbf{a}_{l}^{p}=\operatorname{CA}_{p}(\mathbf{d}_{l},\mathbf{H}_{u}^{p},\mathbf{H}_{u}^{p}),
\qquad
\mathbf{a}_{l}^{g}=\operatorname{CA}_{g}(\mathbf{d}_{l},\mathbf{H}_{u}^{g},\mathbf{H}_{u}^{g}).
\end{equation}
Keeping the operators separate prevents population evidence from being mixed into Personal Memory before the decoder can assess its usefulness. A vector-valued gate adaptively activates Group Memory at each decoding level:
\begin{equation}
\mathbf{g}_{l}=\sigma\!\left(\mathbf{W}_{g}
[\mathbf{d}_{l};\mathbf{a}_{l}^{p};\mathbf{a}_{l}^{g}]+\mathbf{b}_{g}\right),
\qquad
\widetilde{\mathbf{a}}_{l}=\mathbf{a}_{l}^{p}+\mathbf{g}_{l}\odot\mathbf{a}_{l}^{g}.
\end{equation}
This asymmetric residual form keeps Personal Memory as the primary signal and activates Group Memory only where it is compatible with the current decoder state; noisy or weakly supported retrieval can therefore be suppressed without discarding individual context. Group Memory is encoded once per request and reused across decoding steps and expanded beams. Together, direct reasoning injection and dual-memory augmentation play complementary roles: the former supplies a Personal-Memory-conditioned semantic prior at the decoder input, while the latter grounds inferred intent in population behavior and exposes that evidence as auxiliary memory. Neither path increases the autoregressive SID decoding length or beam-search space.

\subsection{Training Recipe}
\label{sec:reason_training}

\subsubsection{Think-Model Training}
\label{sec:reason_training_think}

\paragraph{Offline Think model.} Training the Think model proceeds in three stages that share one backbone. A vocabulary-expansion stage first adds the semantic-ID tokens to the backbone so that items become first-class symbols it can emit. A content-alignment stage then adapts the backbone to the target surface, tuning a small set of trainable tokens so that the model's language is grounded in the platform's items and their descriptions. A reasoning stage finally applies LatentRec: it trains only the LoRA adapters and the Continuous Thought Module under the per-step reasoning loss of \S\ref{sec:reason_latent}, with $K_{\text{train}}{=}2$, loss weight $\lambda{=}0.1$, and temperatures $T_b{=}1$ and $T_s{=}5$, while the backbone stays frozen. Keeping the backbone frozen confines the reasoning signal to a light adapter, which is what lets the module be dropped at $K{=}0$ serving without disturbing the aligned backbone. The chain-of-thought-distilled initialization of \S\ref{sec:reason_latent} sits between alignment and reasoning and is used only when deeper rollout is required.

\subsubsection{Gen-Model Training}

The downstream TGR-GenRec model is trained with the same next-token-prediction objective used by the production SID generator. For each training request, the Think model's top-$K_r$ reason tokens are precomputed and treated as fixed conditioning features. DRI organizes their codewords by SID level, conditions them on Personal Memory, and inserts the resulting representations before the corresponding target codewords. Under teacher forcing, the generator predicts the target SID $\mathbf{y}=(y^1,\ldots,y^L)$ one level at a time:
\begin{equation}
\mathcal{L}_{\mathrm{gen}}
=-\sum_{l=1}^{L} w_l
\log p_{\theta}\!\left(y^l\mid y^{<l},\mathbf{H}_u^{p},\mathcal{R}_u\right),
\end{equation}
where $w_l$ assigns larger weights to coarser SID levels. Gradients update the generator and DRI parameters, but do not propagate into the nearline Think model.

When GMR is enabled, its retrieved context $\mathcal{C}_u$ is added to the conditioning variables, and the Group Encoder and adaptive fusion gate are optimized jointly through the same NTP loss. GMR itself is non-parametric and receives no gradient. No additional reasoning-specific objective is introduced on the Gen-Model side; reason tokens and dual-memory context improve generation by conditioning the standard SID prediction task.

\subsubsection{Offline Generation at Scale}
\label{sec:reason_offline_gen}

Treating reasoning as an offline feature is practical only if reason tokens can be regenerated for the whole user base on a routine schedule and within a fixed compute budget. Two design choices make this possible. A two-pipeline refresh decouples how often the Think model is retrained from how often its predictions are produced, and a batched decoding scheme runs the per-user top-$K_r$ generation cheaply on an unmodified inference engine.

\paragraph{Routine production.} The refresh runs on two cadences that serve different purposes. A weekly pipeline retrains the Think model end to end so that vocabulary expansion and content alignment keep tracking new items and shifting user interests, and it produces one versioned reasoning model each week. A daily pipeline then applies the current model to the users a staleness detector marks for refresh and writes their top-$K_r$ reason tokens into the Reason Store that the generator reads. Separating the two cadences keeps the expensive retraining weekly while the lighter prediction pass stays daily, so freshness and cost can be tuned independently. Retraining on a fixed-size weekly sample rather than the full log holds each cycle within budget without sacrificing the downstream gain, and every weekly model is versioned so that the generator always consumes a complete and self-consistent release rather than a partially updated one.

\paragraph{Offline decoding at scale.} The daily pass is the cost bottleneck, since it must emit per-user top-$K_r$ semantic IDs for tens of millions of users, which amounts to a constrained top-$K_r$ beam search over a short fixed-length code. Neither native path of a high-throughput inference engine (vLLM) fits this shape. The sampling path, even behind a trie logits filter, replaces beam enumeration with stochastic draws and loses top-$K_r$ recall, dropping Hit@5 well below the reference decoder. The native beam-search path preserves validity but runs as an unoptimized loop that decodes slower than the eager baseline, which removes the very speedup that motivated the migration. The mismatch is structural, because a semantic ID is a short code drawn from a constrained vocabulary, exactly the regime that the engine's throughput machinery is not built to enumerate.

Rather than modify the engine, we recast beam search as a fixed number of batched single-token rounds that its scheduler already runs well. A semantic ID spans a fixed number of codebook levels, so decoding completes in that many rounds. Each round expands every live beam by its top-$B$ next tokens in one batched forward pass and prunes back to the $B$ best partial SIDs by cumulative log-probability, which is the same deterministic enumeration as standard beam search rather than an approximation. Continuous batching packs the beams of all users in a cycle into each pass, and prefix caching stores the long shared user-context prefix once, so the marginal cost of a round is a single incremental decoding step. The SID trie serves as a final validity check instead of a per-step callback, since the enumerated beams are already valid in practice, which keeps the fast path free of per-token masking. The speedup therefore comes entirely from the engine's batching and cache reuse under this decoding schedule, not from any change to its kernels or scheduler.

Because the rounds reproduce the reference enumeration exactly, the schedule matches the reference decoder's Hit@5 bucket by bucket at the same beam width, with the cold, warm, and hot buckets all matching or slightly exceeding the baseline, and it emits only valid semantic IDs. The speedup thus carries no accuracy cost. On a production cluster of eight A100 GPUs it shortens the end-to-end generation cycle by $2.2\times$, from $174$ to $80$ minutes, which brings a ten-million-user refresh from about $2.9$ days down to $1.3$. That headroom is what lets the daily pipeline run within budget and keeps reasoning an offline-amortized feature rather than a per-request cost.

\subsection{Deployment}
\label{sec:reason_deployment}
\paragraph{Asynchronous Reason Materialization.} To avoid placing expensive Think-model inference on the critical serving path, TGR-Reason materializes user-level reasoning outputs as reusable \emph{Reason records} in the Reason Store. Request-triggered nearline jobs refresh selected users under gain, freshness, and budget constraints, while scheduled offline jobs backfill missing, stale, failed, or version-incompatible records. Both paths update the store conditionally, ensuring that newly materialized records are consumed only by subsequent requests. Throughout this section the architecture is described in its full design, including the GMR and \texttt{cot\_emb} paths; the deployed default configuration behind the results of \S\ref{sec:reason_results} injects the semantic-ID reason tokens through DRI alone (\S\ref{sec:reason_inject}).

\subsubsection{Serving Efficiency}
\label{sec:reason_deploy_efficiency}
\paragraph{Moving reasoning off the critical path.} TGR-Reason decouples expensive user-level reasoning from latency-critical candidate generation. The Think model may contain billions of parameters and require substantially more computation than the online generator. Running it synchronously for every request would incur prohibitive latency and GPU cost. TGR-Reason instead materializes its outputs as reusable user-level Reason records, keeping the Think model outside the synchronous request path and avoiding new reasoning-trace generation for the current response.

At serving time, the online generator reads the most recent valid Reason record through a key--value point lookup. The pre-materialized DRI representation is injected into the decoder side of TGR-GenRec as a fixed-size soft prompt. In parallel, the stored complete reason SID tokens are used by GMR to retrieve a bounded set of group-level behavioral memories. The retrieved memories are pooled and incorporated into the context-encoding stage before SID decoding. DRI and GMR thus provide complementary conditioning signals: DRI transfers user-specific reasoning intent directly to generation, whereas GMR grounds generation in population-level behavioral evidence.

The stored SID predictions additionally support a decoder-free direct retrieval path. Specifically, \texttt{topk\_sid} is resolved through the SID index to obtain exact item candidates, after which invalid, unavailable, or out-of-domain items are filtered. Unlike DRI and GMR, which condition TGR-GenRec, direct retrieval directly produces candidate items and does not participate in autoregressive decoding. The generated and directly retrieved candidates are assigned independent quotas before being returned to the downstream ranking stage.

DRI injection, GMR retrieval and pooling, and SID-index lookup are all bounded non-autoregressive operations. They add neither new reasoning-token generation nor additional SID decoding steps, and therefore leave the SID decoding horizon, beam width, and beam-search space unchanged relative to the original TGR-GenRec serving path. The additional request-time cost is limited to Reason-Store access, metadata validation, fixed-size feature projection, bounded Group-Memory retrieval and aggregation, and SID resolution, rather than another round of Think-model reasoning. The expensive reasoning cost is instead amortized over subsequent requests through budget-controlled nearline refreshes and offline backfills.

\subsubsection{Online Serving}
\label{sec:reason_deploy_serving}

\paragraph{Part A: online candidate generation and ranking.} Upon receiving a user request, the online service retrieves the user's conventional sequence, profile, and contextual features. In parallel, it performs a point lookup in the Reason Store using the user identifier. A Reason record is considered viable only if it is present, unexpired, supported by the deployed model and feature schema, and associated with the active SID-codebook version.

When the record is viable, its DRI representation and reason SID tokens are consumed by TGR-GenRec through complementary conditioning paths. The DRI representation is injected into the decoder as a fixed-size soft prompt, while GMR uses the complete reason SID tokens to retrieve group-level memory and incorporates the resulting memory representation into the context encoder before SID decoding. TGR-GenRec then autoregressively generates a SID list. In parallel, the direct retrieval path performs an exact lookup of the stored \texttt{topk\_sid}. The SID index maps the predicted SIDs to item identifiers and applies validity and availability filters. Candidate items from the generative and direct retrieval paths are assigned separate source tags and quotas before being passed to the downstream ranking stage.

Reason features are also reused by the ranking model. In particular, the valid Reason record is projected into rank-side features and provided to CCFormer in TGR-GenRank, complementing the conventional user, item, and context features. Thus, the same materialized Reason record improves both candidate generation and downstream ranking without requiring synchronous Think-model inference.

A Reason-store miss or an invalid record never blocks the request. TGR-GenRec falls back to its original no-Reason generation path, the direct retrieval path is skipped and its quota is filled by baseline candidate sources, and TGR-GenRank uses its base CCFormer feature set without Reason-derived features. Thus, the availability of the online recommendation service does not depend on successful execution of the Think model.

\paragraph{Part B: request-triggered nearline refresh.} Every incoming request may asynchronously emit a lightweight trigger event, such as \texttt{user\_id}, request time, and scene identifier, to the Trigger Event Queue. This event emission follows a fire-and-forget pattern and proceeds in parallel with the online candidate-generation path; the current request neither waits for nor consumes the newly generated Reason record.

The RTC consumes these events and determines whether recomputing the user's Reason record is worth the associated inference cost. Its admission policy jointly considers:
\begin{enumerate}
    \item whether a compatible Reason record is missing or approaching
          expiration;
    \item the degree of user-behavior change since the previous
          \texttt{behavior\_cutoff};
    \item the expected marginal recommendation gain for the current user
          or user segment;
    \item user-level cooldown and single-flight constraints; 
    \item the current inference budget, queue backlog, and GPU capacity.
\end{enumerate}
Only admitted jobs are submitted to the resident-GPU TGR-Reason serving pool. The pool uses dynamic micro-batching to improve utilization while maintaining second-level freshness. Its output is validated and conditionally written to the Reason Store. The refreshed record becomes visible to subsequent requests rather than retroactively affecting the request that triggered the update.

\paragraph{Part C: offline backfill and model update.} The offline tier operates at user granularity and prioritizes throughput rather than individual-job latency. A scheduled stale-user detector scans the user log and Reason-store metadata to identify:
\begin{itemize}
    \item users with no materialized Reason record;
    \item users whose behavior has advanced beyond the stored
          \texttt{behavior\_cutoff};
    \item expired or low-quality records;
    \item failed or skipped nearline refreshes; 
    \item records produced by an obsolete Think-model, prompt, feature,
          or SID-codebook version.
\end{itemize}
The selected users are processed by an elastic large-batch reasoning pool, which maximizes GPU throughput and writes the regenerated records through the same conditional-update interface used by the nearline tier. In our deployment, this backfill is executed daily, whereas model retraining and checkpoint publication follow a lower-frequency schedule, e.g., weekly. Publishing a new model version marks incompatible Reason records as stale and triggers their progressive reconstruction through the nearline and offline refresh mechanisms.

\subsubsection{Store Consistency and Failure Isolation}
\label{sec:reason_deploy_consistency}

\paragraph{Versioned Reason records.} The Reason Store is a distributed key--value abstraction backed by a managed key--value service on Tencent Cloud, rather than being tied to a particular storage engine. Each record is keyed by the user identifier, optionally augmented with the recommendation scene and experiment variant, and contains
\begin{equation}
\begin{aligned}
\{&
    \texttt{cot\_emb},
    \texttt{topk\_sid},
    \texttt{sid\_emb},
    \texttt{behavior\_cutoff}, \\
  & \texttt{generated\_at},
    \texttt{model\_version},
    \texttt{sid\_codebook\_version},
    \texttt{expire\_at}
\}.
\end{aligned}
\end{equation}
The underlying managed service provides low-latency point lookup, TTL-based expiration, and atomic conditional-update semantics. This design keeps the serving interface independent of a specific storage product while allowing the storage tier to be selected according to access density, capacity, latency, and cost requirements.

\paragraph{Conditional updates.} Nearline and offline jobs may finish out of order. For example, an offline job based on user behavior observed up to 18:00 may complete after a nearline job based on behavior observed up to 20:00. Comparing only \texttt{generated\_at} would incorrectly allow the later-finishing but semantically older record to overwrite the fresher one.

TGR-Reason therefore accepts only records compatible with the active model and feature versions. Within the same version, an incoming record may replace the stored record only when its \texttt{behavior\_cutoff} is newer. The \texttt{generated\_at} field is retained for monitoring and auditing, but is not used as the primary freshness order. This version-aware conditional-write policy prevents stale offline jobs from overwriting newer nearline results.

\paragraph{Failure isolation.} The three tiers are deliberately failure-isolated. A failure in the trigger queue, RTC, or nearline serving pool only reduces Reason freshness. A failure in the offline batch pool only delays coverage recovery. Neither failure mode interrupts the online request path, which can continue with the most recent compatible Reason record or fall back to no-Reason TGR-GenRec. This separation allows the system to trade reasoning freshness against inference budget without coupling large-model availability to the availability or tail latency of online recommendation.

\subsection{Evaluation Results}
\label{sec:reason_results}

\paragraph{Offline results.} Table~\ref{tab:reason_offline} compares TGR-Reason with the deployed production generator on one of Tencent's commercial content platforms. We report Hit@$K$ at five cutoffs over three progressively more focused cohorts: all recommendation traffic, primary recommendation traffic, and cold-start users within the primary recommendation traffic. TGR-Reason improves every cutoff in every cohort, showing that the reasoning signal is useful not only for the top-ranked prediction but throughout the retrieved candidate set.

The improvement is strongest near the head of the list. Across all recommendation traffic, Hit@1 rises from $0.3385$ to $0.4339$ ($+28.2\%$), while Hit@5 and Hit@50 improve by $10.4\%$ and $2.4\%$, respectively. Restricting evaluation to the primary recommendation traffic increases the Hit@1 gain to $58.6\%$ ($0.2247\rightarrow0.3563$), showing that the benefit of reason-token injection is especially pronounced in the platform's primary recommendation traffic. The gain narrows as the cutoff $K$ grows because the production model already recovers many relevant items in a wider candidate set; reasoning contributes most by moving a plausible target toward the first few positions.

Cold-start users within the primary recommendation traffic exhibit the largest improvement. Their Hit@1 increases from $0.0451$ to $0.2606$, an absolute gain of $0.2155$ and a relative gain of $477.8\%$. The effect remains substantial at larger cutoffs: Hit@5 improves by $39.2\%$, Hit@10 by $24.9\%$, Hit@20 by $20.1\%$, and Hit@50 by $11.6\%$. This pattern supports the central motivation of TGR-Reason: when behavioral history is sparse, a co-occurrence-driven generator has little evidence with which to resolve intent, whereas reasoning-enriched tokens provide a semantic prior that narrows the target space. As user history and candidate-set width increase, collaborative evidence recovers and the marginal contribution of that prior naturally becomes smaller.

\begin{table}[t]
\centering
\caption{Offline comparison between reason-token injection and the online TGR-GenRec baseline on one of Tencent's commercial content platforms. Results are averaged with equal weight over three test dates and use the deployed default configuration (Personal Encoder $+$ Direct Reasoning Injection). The GMR branch is excluded. Relative gains are shown in italics.}

\label{tab:reason_offline}
\small
\setlength{\tabcolsep}{5pt}
\begin{tabular}{l|l|ccccc}
\toprule
Cohort & Model & Hit@1 & Hit@5 & Hit@10 & Hit@20 & Hit@50 \\
\midrule
\multirow{3}{*}{All rec.\ traffic}
 & TGR-GenRec (online) & 0.3385 & 0.5152 & 0.5792 & 0.6377 & 0.7063 \\
 & \textbf{TGR-Reason} & \textbf{0.4339} & \textbf{0.5686} & \textbf{0.6155} & \textbf{0.6635} & \textbf{0.7233} \\
 & \emph{Rel.\ gain} & \emph{+28.2\%} & \emph{+10.4\%} & \emph{+6.3\%} & \emph{+4.1\%} & \emph{+2.4\%} \\
\midrule
\multirow{3}{*}{Primary rec.\ traffic}
 & TGR-GenRec (online) & 0.2247 & 0.4306 & 0.5116 & 0.5859 & 0.6671 \\
 & \textbf{TGR-Reason} & \textbf{0.3563} & \textbf{0.4870} & \textbf{0.5461} & \textbf{0.6085} & \textbf{0.6847} \\
 & \emph{Rel.\ gain} & \emph{+58.6\%} & \emph{+13.1\%} & \emph{+6.7\%} & \emph{+3.9\%} & \emph{+2.6\%} \\
\midrule
\multirow{3}{*}{\shortstack[l]{Primary rec.,\\\textbf{cold-start}}}
 & TGR-GenRec (online) & 0.0451 & 0.3379 & 0.4551 & 0.5570 & 0.6949 \\
 & \textbf{TGR-Reason} & \textbf{0.2606} & \textbf{0.4702} & \textbf{0.5684} & \textbf{0.6687} & \textbf{0.7753} \\
 & \emph{Rel.\ gain} & \textbf{\emph{+477.8\%}} & \emph{+39.2\%} & \emph{+24.9\%} & \emph{+20.1\%} & \emph{+11.6\%} \\
\bottomrule
\end{tabular}
\end{table}

\paragraph{Online A/B.} We further evaluate TGR-Reason in a live online A/B experiment on the same Tencent commercial content platform, using the incumbent production generator as the control. As shown in Table~\ref{tab:reason_online}, TGR-Reason improves the Effective Consumption Rate by $1.75\%$ and the Exposure-to-Conversion Rate for new users by $13.09\%$; both lifts are statistically significant. The first result shows that the offline gain translates into more exposures producing meaningful content consumption rather than merely increasing candidate recall. The substantially larger new-user conversion lift is also consistent with the offline cohort analysis: reasoning-enriched SIDs supply useful semantic evidence precisely where personal interaction history is least informative. Together, the online results confirm that the cold-start improvements in Table~\ref{tab:reason_offline} survive the complete production cascade and produce measurable user-value gains.

\begin{table}[t]
\centering
\caption{Online A/B results of TGR-Reason on one of Tencent's commercial content platforms. Relative lifts are measured against the incumbent production generator; both improvements are statistically significant.}
\label{tab:reason_online}
\small
\setlength{\tabcolsep}{6pt}
\begin{tabular}{lcc}
\toprule
Metric & Relative lift & Significance \\
\midrule
Effective Consumption Rate & $+1.75\%$ & Significant \\
Exposure-to-Conversion Rate (New Users) & $+13.09\%$ & Significant \\
\bottomrule
\end{tabular}
\end{table}

\section{Conclusion, Limitations, and Future Directions}
\label{sec:conclusion}

\subsection{Conclusion}

In this report we introduced \textbf{TGR}, an industrial-scale generative recommendation stack consisting of three coupled directions: TGR-GenRank
 (generative-paradigm ranking), TGR-GenRec (unified end-to-end generative recommendation), and TGR-Reason (reasoning-augmented generative recommendation). The three directions rest on
shared foundations: the generative-recommendation models
share a semantic-ID tokenizer family (PCRQ-VAE for slate generation, OSQ-VAE
for NTP generation) and a common user/context encoding interface; the whole
stack is trained and served on the Numerous-Torch infrastructure. TGR-GenRank is realized by CCFormer, which unifies feature-field separated cross attention with subspace token mixing and hierarchical sequence compression: it consistently outperforms strong industrial baselines offline, follows predictable scaling laws in sequence length and model size at roughly half the GFLOPs of HSTU, trains $2.21\times$ faster than HSTU, and is fully deployed in a video-recommendation scenario ($+3.57\%$ CTR) and an advertising-ranking scenario ($+1.71\%$ advertising revenue). The most fully developed component, TGR-GenRec, spans one production model per generation paradigm: BARGE, which structurally adapts next-token-prediction SID generation through item context-aware attention, hierarchical path reranking, and dual-path decoding at zero additional serving cost; and HiGR, which realizes whole-slate generation by combining Prefix-Contrastive RQ-VAE, a Hierarchical Slate Decoder, and ORPO-based listwise multi-objective preference alignment. TGR-Reason is realized by a LatentRec-trained Think model whose reason tokens, exported offline, are injected into the production generator through Direct Reasoning Injection at zero request-path reasoning cost, raising cold-start new-user Hit@1 by $+477.8\%$ and delivering statistically significant online gains ($+1.75\%$ Effective Consumption Rate, $+13.09\%$ new-user Exposure-to-Conversion Rate). TGR has been deployed on multiple Tencent surfaces, serving hundreds of millions of users.

\subsection{Limitations}

The limitations below reflect where the stack stands today.

\begin{itemize}[leftmargin=1.2em,itemsep=1pt,topsep=2pt]
\item \textbf{A stack of coupled models, not yet one model.} TGR upgrades or replaces the cascade piece by piece: TGR-GenRank still emits per-item multi-task scores that the downstream cascade consumes, and TGR-GenRec currently ships two production models with two tokenizers (OSQ-VAE for BARGE, PCRQ-VAE for HiGR). 
\item \textbf{Cold-start items.} The hierarchical SID space inherits the cold-start limitations of vector-quantized tokenizers: a brand-new item enters the code space on semantic evidence alone, before the collaborative relations that shape SID prefixes (\S\ref{sec:higr_pcrqvae}) can be estimated for it. TGR-Reason's demonstrated cold-start gains concern new \emph{users} (\S\ref{sec:think}); cold-start \emph{items} remain governed by tokenizer quality.
\item \textbf{Global alignment trade-offs.} ORPO-based alignment integrates ranking fidelity, genuine user interest, and diversity under a single global weighting (\S\ref{sec:higr-method}), uniform across users and contexts; a personalized or context-conditioned weighting of the three listwise objectives remains open.
\item \textbf{Reasoning is amortized, not deliberative.} TGR-Reason spends all reasoning off the request path: the Think model is retrained weekly, reason tokens are regenerated daily (and, for trigger-gated users, within seconds in the nearline tier), and serving reads tokens precomputed at $K{=}0$. Intra-session intent shifts therefore reach the generator only through its own input context, and explicit in-text reasoning remains too expensive for the request path.
\end{itemize}

\subsection{Future Directions}

\begin{enumerate}[leftmargin=1.4em,itemsep=2pt,topsep=2pt]
\item \textbf{Bending the scaling curve, not just riding it.} The scaling frontier of generative ranking has moved from demonstrating scaling laws \citep{hstu_2024} to improving their slope: ULTRA-HSTU reports over $5\times$ faster training scaling and $21\times$ faster inference scaling through end-to-end model--system co-design \citep{ultrahstu_2026}. TGR will push HiGR beyond the $0.05$B--$2$B range in which its scaling law has been observed (\S\ref{sec:higr-offline})---including a unified lazy-decoder realization of planner and generator---and pursue the same co-design discipline (sparse attention, input-sequence and topology design) on Numerous-Torch, always inside the $<50$\,ms P99, fixed-GPU envelope that keeps TGR-GenRec deployable (\S\ref{sec:higr-efficiency}).
\item \textbf{Reasoning-native generation and test-time compute.} Reasoning is becoming a first-class citizen of industrial recommendation rather than an offline add-on: OneRec-Think brings in-text chain-of-thought into the generator \citep{onerec_think_2025}, ReaRec explores latent multi-step reasoning at inference \citep{rearec_2025}, and RecGPT-V3 internalizes verbose rationales into compact latent tokens backed by a stateful memory hub \citep{recgpt_v3_2026}---independent convergence on the latent, amortized route that TGR-Reason takes (\S\ref{sec:think}). TGR will extend this route with selective request-time deliberation ($K{>}0$ decoding for high-value or cold-start requests), distillation into latency-tolerant variants in the spirit of reasoning-augmented LLMs \citep{deepseek_r1_2025}, and richer memory beyond the current Personal/Group split.
\item \textbf{Agentic self-evolution of the recommender.} Self-evolution has moved from vision to production practice: LLM agents that autonomously propose, screen, and A/B-validate improvements to a large-scale industrial recommender---spanning optimizers, architectures, and reward functions---have been shown to surpass conventional engineering workflows in development velocity and model performance \citep{selfevolve_rec_2026}, and surveys chart the same trajectory for agents at large \citep{selfevolve_2025}. TGR will likewise advance along this self-evolution direction in its future iterations.
\end{enumerate}

\appendix
\section*{Appendix}
\addcontentsline{toc}{section}{Appendix}

\section{Contributions}
\label{app:contributions}

\noindent The listing of contributors is in alphabetical order based on their last names.

\medskip
\begin{multicols}{3}
\noindent
Lei Cheng\\
Haonan Hu\\
Beibei Kong\\
Yudong Li\\
Zang Li\\
Yunsheng Pang\\
Hongyang Su\\
Jianchao Tu\\
Yunlong Wang\\
Bing Wen\\
Junzhang Zhu\\
Shaojie Zhu\\
Chengxiang Zhuo
\end{multicols}

\section{Notation}
\label{app:notation}

Table~\ref{tab:notation} summarizes the stack-level notation used throughout the report, followed by the principal model-specific symbols of each technical chapter. Because the three production models are presented as independent, parallel works (\S\ref{sec:higr}) and follow the notation of their source papers, symbols below are scoped to the sections listed, and the same letter may denote different quantities in different chapters---e.g., $L$ is the per-item SID depth in \S\ref{sec:barge} and \S\ref{sec:think} (denoted $D$ elsewhere in the report) but the number of interaction blocks in \S\ref{sec:ccformer}. Symbols not listed here are local to the subsection in which they appear.

\begin{table}[!htbp]
\centering
\caption{Notation. The top block lists stack-level symbols; the lower blocks list the principal chapter-specific symbols, following the corresponding source papers.}
\label{tab:notation}
\small
\setlength{\tabcolsep}{5pt}
\begin{tabular}{l|l}
\toprule
Symbol                       & Meaning \\
\midrule
\multicolumn{2}{l}{\emph{Stack-level (used throughout)}} \\
$u, \mathcal{U}$             & User; set of users \\
$v, \mathcal{V}$             & Item; set of items \\
$S^u$                        & Chronological interaction history of user $u$ \\
$O^u$                        & Output slate of $M$ items for user $u$ \\
$M$                          & Number of items in a slate \\
$D$                          & Number of SID levels per item \\
$s^d, s^d_m$                 & $d$-th SID; $d$-th SID of item $m$ \\
$B$                          & Beam width \\
$d_{\text{model}}$           & Hidden dimension of the unified encoder and the HiGR decoders \\
$C$                          & User / context representation (encoder output) \\
$\mathcal{F}_\theta$         & Generative model decoding the output slate \\
\midrule
\multicolumn{2}{l}{\emph{TGR-GenRank / CCFormer (\S\ref{sec:ccformer})}} \\
$\mathbf{U}, \mathbf{S}, \mathbf{T}$ & User-profile, behavior-sequence, and target-item token fields \\
$L_u, L_s, L_t$              & Numbers of tokens in the three fields \\
$L$                          & Number of stacked CCFormer interaction blocks \\
$m, n$                       & Sequence and channel group sizes in subspace token mixing \\
$k, s$                       & Kernel size and stride of convolutional sequence compression \\
$\alpha, \beta, \gamma$      & Learnable parameters of the temporal decay $\alpha\,\beta^{|t_i - t_j|^{\gamma}}$ \\
\midrule
\multicolumn{2}{l}{\emph{TGR-GenRec / BARGE (\S\ref{sec:barge})}} \\
$L$                          & Per-item SID depth; $c_l$ is the codeword at layer $l$ \\
$\mathbf{h}_0$               & Pre-generation decoder hidden state (HPR context anchor) \\
$\mathbf{p}^{(l)}$           & Cumulative path embedding $\sum_{j=1}^{l}\mathbf{e}_{c_j}$ \\
$\lambda$; $N$               & HPR reranking weight; HPR scoring-pool size \\
$R$                          & Householder-parameterized orthogonal rotation of OSQ-VAE \\
$K$                          & Length of the final merged recommendation list \\
\midrule
\multicolumn{2}{l}{\emph{TGR-GenRec / HiGR (\S\ref{sec:higr_model})}} \\
$z, \hat z$                  & Latent embedding and aggregated quantized form (PCRQ-VAE) \\
$e^d_a$                      & Codeword embedding of item $a$ at quantization layer $d$ \\
$\eta$                       & Commitment strength in the global-quantization loss \\
$\lambda_1, \lambda_2$       & Weights of the global-quantization and contrastive losses \\
$\tau$; $w_d$                & InfoNCE temperature; per-layer weight in the prefix contrastive loss \\
$\hat i_m$                   & Planned preference embedding for slate position $m$ (HSD) \\
$l_{\text{slate}}, l_{\text{item}}$ & Layer counts of the slate planner and item generator \\
$y^+, y^-$                   & Positive / negative slate in an ORPO preference pair \\
$\alpha$; $\beta$            & ORPO alignment coefficient; ILD regularization weight \\
\midrule
\multicolumn{2}{l}{\emph{TGR-Reason (\S\ref{sec:think})}} \\
$K$                          & Latent reasoning steps of LatentRec ($K_{\text{train}}$ in training; $K{=}0$ serving) \\
$K_r$                        & Number of exported reason tokens per user \\
$\mathcal{R}_u$              & Reason-token set of user $u$; each $\mathbf{r}_{u,k}$ is a complete $L$-level SID \\
$W$                          & Forward window of Group Memory retrieval \\
$M$                          & Successors retained per reason-token query in GMR \\
$\mathbf{H}_u^{p}, \mathbf{H}_u^{g}$ & Personal-Memory and Group-Memory representations \\
\bottomrule
\end{tabular}
\end{table}

\section{Default Configurations and Hyperparameters}
\label{app:configs}

Table~\ref{tab:configs} consolidates the default configurations of the three production models and TGR-Reason as a single reference, drawn from the corresponding source papers \citep{ccformer_2027,barge_2027,higr_2026} and \S\ref{sec:think}; the body sections describe the context in which each setting is used. Values not published by the source papers (e.g., the ILD regularization weight $\beta$ of HiGR) are omitted.

\begin{table}[!htbp]
\centering
\caption{Default configurations and hyperparameters. CCFormer: industrial setting of \citep{ccformer_2027}; BARGE: public-benchmark setting of \citep{barge_2027} (the deployed retrieval channel is described in \S\ref{sec:barge_deployment}); HiGR: industrial setting of \citep{higr_2026}; TGR-Reason: deployed configuration of \S\ref{sec:think}.}
\label{tab:configs}
\small
\setlength{\tabcolsep}{5pt}
\begin{tabular}{l|l}
\toprule
Parameter & Default value \\
\midrule
\multicolumn{2}{l}{\emph{CCFormer (\S\ref{sec:ccformer})}} \\
Feature-token dimensionality $d$    & 256 \\
Interaction blocks $L$              & 8 \\
Subspace group sizes $(m, n)$       & $(8, 16)$ \\
Compression kernel / stride $(k, s)$ & $(3, 2)$ \\
Optimizer / learning rate / batch   & Adam / $10^{-4}$ / 4096 \\
Behavior-sequence length            & 1000 industrial; 200 public; 500--2000 in scaling \\
Training hardware                   & 16$\times$ NVIDIA H20 \\
Precision and sparse compression    & BF16/FP16; INT8 ($\sim$70\%) $+$ double hashing ($\sim$50\%) \\
Hyperparameter search               & TPE, applied to every compared method \\
\midrule
\multicolumn{2}{l}{\emph{BARGE (\S\ref{sec:barge})}} \\
SID depth $L$ / codebook sizes      & 4 / $(512, 256, 128, 64)$, all layers learned \\
Tokenizer                           & OSQ-VAE: Householder rotation $R$, two channels \\
Encoder / decoder towers            & 2 layers, 4 heads each (shared encoder, dual decoders) \\
Embedding / attention / FFN dims    & 128 / 512 / 1024 \\
Optimizer / batch / epochs          & Adam with warmup / 256 / 200 (early stopping; 3 seeds) \\
Beam width $B$ / HPR pool $N$       & 20 / 400 \\
HPR fusion weight $\lambda$         & 0.25 \\
OR-fusion operator                  & LSE (log-sum-exp soft OR) \\
Training hardware                   & 2$\times$ NVIDIA H20 \\
\midrule
\multicolumn{2}{l}{\emph{HiGR (\S\ref{sec:higr_model})}} \\
Hidden / FFN dimensions             & $d_{\text{model}} = 512$ / $d_{\text{FFN}} = 2048$ \\
SID depth $D$ / codebook size       & 3 / 1024 per layer \\
Planner / generator depth           & $l_{\text{slate}} = 14$ / $l_{\text{item}} = 2$ \\
PCRQ-VAE weights                    & $\eta = 0.1$, $\lambda_1 = 0.1$, $\lambda_2 = 0.01$, $(w_1, w_2) = (1, 0.1)$ \\
ORPO coefficient $\alpha$           & 0.1 \\
Training data                       & 1B pre-training samples; 3\% for ORPO post-training \\
Model scale                         & 25M / 100M in deployed studies; 0.05B--2B in scaling \\
Hardware and serving                & NVIDIA H20 training; L20 serving with KV caching \\
\midrule
\multicolumn{2}{l}{\emph{TGR-Reason (\S\ref{sec:think})}} \\
Think backbone                      & Qwen3-1.7B $+$ LoRA (backbone frozen in the reasoning stage) \\
Latent reasoning steps              & $K_{\text{train}} = 2$; $K = 0$ at offline generation \\
PRL weight / temperatures           & $\lambda = 0.1$; $T_b = 1$, $T_s = 5$ \\
Soft tokens $N$                     & 64 ($\sim$2.5M trainable STS parameters) \\
Offline decoding                    & vLLM batched rounds on 8$\times$ A100; $2.2\times$ cycle speedup \\
\bottomrule
\end{tabular}
\end{table}

\section{Complexity Analysis}
\label{app:complexity}

\paragraph{CCFormer (\S\ref{sec:ccformer}).}
CCFormer avoids the $\mathcal{O}(L_s^2)$ cost of global self-attention over the behavior sequence by construction: relative temporal-positional encoding is applied within local groups of $m$ behaviors at $\mathcal{O}(L_s m)$ cost, subspace token mixing processes the sequence at cost linear in $L_s$, and the strided convolution after each block hands a geometrically shorter sequence to the next block, so deeper blocks retain access to the full history at a fraction of its length. At matched capacity this yields roughly half the per-sample GFLOPs of HSTU (Table~\ref{tab:ccformer_size}).

\paragraph{BARGE (\S\ref{sec:barge}).}
In hierarchical SID beam search, inference latency scales approximately linearly with the beam width $B$ and decoding memory grows with $B \times L$, which is why BARGE corrects semantic drift at a fixed $B$ rather than by widening the beam. HPR adds only a lightweight per-layer dual-tower scoring over a pool of $N$ candidates, with the context-side projection computed once per request; DPD runs its two decoder towers in parallel on a shared encoder while keeping the per-channel beam width and the final Top-$K$ list length unchanged. The combined parameter count remains below the TIGER baseline (Table~\ref{tab:barge_efficiency}).

\paragraph{HiGR vs.\ OneRec-style decoding (\S\ref{sec:higr_model}).}
For a slate of $M$ items, $D$ SIDs per item, hidden size $d$, beam width $B$, $l_{\text{slate}}$ slate-planner layers (equal to the lazy-decoder depth of the OneRec-style baseline), and $l_{\text{item}}$ item-generator layers, decomposing full-sequence beam decoding into greedy slate planning followed by local per-item beam search reduces the decoding complexity from
\[
\mathcal{O}\!\big(B \cdot M^3 D^3 \cdot l_{\text{slate}} \cdot d\big) \quad \text{(OneRec-style)}
\quad \longrightarrow \quad
\mathcal{O}\!\big(M^3 \cdot l_{\text{slate}} \cdot d \;+\; B \cdot M D^3 \cdot l_{\text{item}} \cdot d\big) \quad \text{(HiGR)},
\]
which underlies the layer-allocation and serving results of \S\ref{sec:higr-efficiency}.

\clearpage

\end{document}